\documentclass[sigconf,nonacm]{acmart}
\AtBeginDocument{%
  \providecommand\BibTeX{{%
    \normalfont B\kern-0.5em{\scshape i\kern-0.25em b}\kern-0.8em\TeX}}}

\setcopyright{none}

\renewcommand\footnotetextcopyrightpermission[1]{}

\usepackage{xcolor, soul}
\usepackage{color}
\usepackage{graphics}
\usepackage{graphicx}
\usepackage{url}
\usepackage{multirow}
\usepackage{xspace}
\usepackage{enumitem}
\usepackage{subcaption}
\usepackage{balance}

\usepackage{booktabs}
\usepackage{colortbl}
\usepackage[normalem]{ulem}

\definecolor{lvlC}{HTML}{2A78D6}
\definecolor{lvlR}{HTML}{EB6834}
\definecolor{lvlI}{HTML}{1BAF7A}
\definecolor{rowband}{HTML}{F1F5F9}
\definecolor{arrowgray}{HTML}{94A3B8}

\newcommand{\myparatight}[1]{\vspace{0.25ex}\noindent\textbf{#1~~}}

\newcommand{\mycode}[1]{\texttt{#1}}
\newcommand{\myLevelOne}{\mycode{L1-C}}
\newcommand{\modellogo}[1]{\raisebox{-0.17em}{\includegraphics[height=0.95em]{figs/logos/#1}}\hspace{0.35em}}
\newcommand{\myLevelTwo}{\mycode{L2-R}}
\newcommand{\myLevelThree}{\mycode{L3-I}}

\newif\ifshowcomments
\showcommentstrue

\usepackage{xspace}
\newcommand{\labname}{\sc EmbedGym}
\newcommand{\labnamebf}{\sc\textbf{EmbedGym}}
\newcommand{\benchname}{\sc EmbedGym-Bench}
\newcommand{\benchnamebf}{\sc\textbf{EmbedGym-Bench}}
\newcommand{\labbenchnamebf}{\sc\textbf{EmbedGym(-Bench)}}

\newcommand{\remove}[1]{}

\newcommand{\del}[1]{}

\begin{document}
\brokenpenalty=0 %
\title[]{{\labnamebf}: Letting Coding Agents Develop and Debug Embedded Firmware Without Physical Hardware}

\author{Minghui Zhao, Yiming Li, Sophie Zhu, Maxwell Khor, Lauren Ouyang$^{\ast}$, \texorpdfstring{\\}{ } Hai ``Helen'' Li, Yiran Chen, and Tingjun Chen}
\affiliation{%
  \institution{Department of Electrical and Computer Engineering, Duke University;~ $^{\ast}$Tesla STEM High School \vspace{0.5ex}}
  \country{}}
\affiliation{%
  \institution{\LARGE {\urlstyle{rm}\url{https://embedgym.github.io}}}
  \country{}}

\renewcommand{\shortauthors}{Zhao et al.}

\begin{abstract}
Modern coding agents perform well on general software benchmarks because they can run code and unit tests in a closed loop.
Embedded firmware development, however, breaks this loop: a program that compiles can still fail on real hardware due to pin polarity, bus protocol, timing, or networking bugs that an agent at a text terminal cannot observe.
Existing embedded coding agents either skip hardware execution, drive GPIO-only simulators with multi-second delays, or test on physical boards in a static environment.
In this paper, we present {\labname}, an interactive virtual hardware environment that lets coding agents develop and debug embedded firmware without physical hardware.
{\labname} uses instruction-level microcontroller unit (MCU) emulators to execute the same binary that would be flashed onto a physical board, connects them to 41 peripheral models, an analog circuit simulator, and virtual WiFi/Bluetooth Low Energy (BLE) endpoints, and exposes an interface for agents to change sensor inputs, press buttons, and inspect hardware states.
On top of it, we build {\benchname}, which consists of 598 tasks across 6 MCU-framework pairs graded by hidden timed test vectors, using a five-stage multi-agent workflow that synthesizes peripheral models and tasks from datasheets and driver libraries.
We expose {\labname} to a coding agent at three levels and evaluate eight models over more than 14K trajectories.
Even when compilation succeeds 99.7\% of the time, 19.9\% of compiled firmwares fail at runtime (up to 73\% on individual targets).
In contrast, interactive hardware testing in {\labname} increases pass rates across models by $+$4.9\% to $+$14.7\% (up to $+$31\% on individual targets).
Our results also show that an agent with {\labname} outperforms prior embedded agent frameworks, and lightweight models with {\labname} match frontier models without it at 30--70$\times$ lower cost.
\end{abstract}

\maketitle

\section{Introduction} \label{sec:intro}

\begin{figure*}[!t]
\centering
\includegraphics[width=0.99\linewidth]{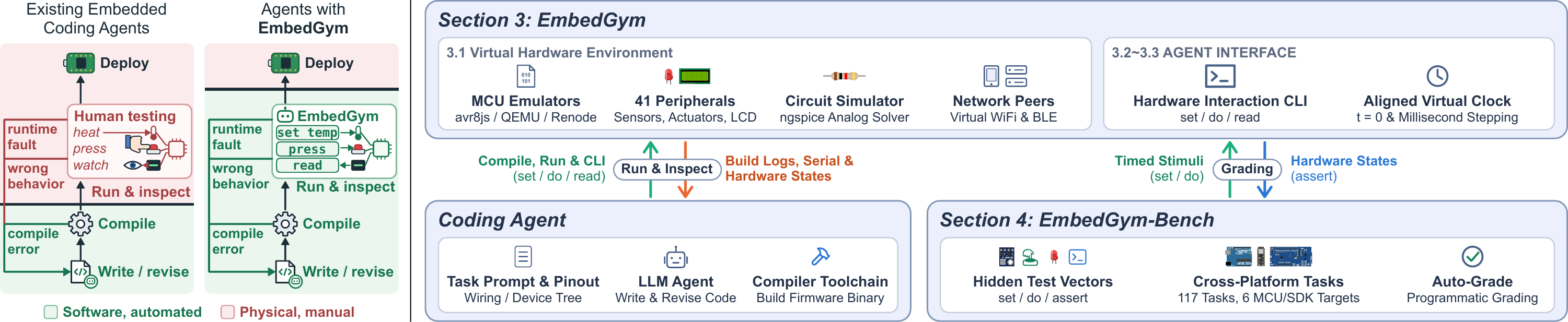}
\vspace{-2.0mm}
\caption{Overview of {\labnamebf} and {\benchnamebf}: a coding agent iteratively writes, compiles, and tests embedded firmware through {\labnamebf}'s command-line interface against instruction-level MCU emulators, 41 peripheral models, an \texttt{ngspice} circuit solver, and virtual WiFi/BLE endpoints (\S\ref{sec:system}), while {\benchnamebf} evaluates the submitted binary across 598 tasks over 6 MCU-framework pairs using hidden, precisely timed test vectors (\S\ref{sec:workflow}).}
\label{fig:overview}
\vspace{-3.0mm}
\end{figure*}

Large language models (LLMs) and coding agents have achieved strong performance on general software engineering benchmarks such as SWE-bench~\cite{jimenez2024swe}, Terminal-Bench~\cite{merrill2026terminal}, FrontierCode~\cite{frontiercode2026}, and CursorBench~\cite{cursorbench2026}, largely because they can run programs and unit tests directly on the host machine to diagnose and repair bugs in a closed loop.
Developing and verifying embedded firmware, however, is much more challenging.
Embedded systems code is cross-compiled for specific architectures of the microcontroller unit (MCU) and interacts with the physical world by reading and writing hardware registers across external sensors, actuators, analog circuits, and wireless networks. 
As a result, embedded firmware that compiles without errors and can be successfully flashed to the MCU may frequently fail on real hardware due to wrong register bitmasks, inverted pin polarities, incorrect sensor scaling formulas, or protocol timing violations.

Existing embedded coding agents or benchmarks cannot support closed-loop hardware debugging or reliable runtime evaluation for three main reasons. 
\emph{First}, prior benchmarks rely on indirect evaluation criteria rather than verifying true hardware state. 
IoTPilot~\cite{gong2025programming} checks only whether code compiles and contains reference API names in the source text, AutoEmbed~\cite{yang2026autoembed} flashes code onto physical boards and checks LLM-inserted serial print statements, and IoT-Skills\-Bench~\cite{li2026skilled} evaluates one-shot generated firmware on physical boards without closed-loop runtime feedback.
When a board sits in a static environment with constant sensor readings and unpressed buttons, passive serial logs mask input-dependent bugs and fail to guide agent debugging.
\emph{Second}, existing simulator-based approaches lack peripheral coverage and timing precision. EmbedAgent~\cite{xu2026embedagent} drives the browser-based Wokwi~\cite{wokwi} simulator externally, but supports only nine basic GPIO components without bus sensors, analog circuits, or WiFi and Bluetooth Low Energy (BLE) networks.
Driving a browser over host wall-clock time also forces a 2-second hold delay per state change to mask network jitter, preventing millisecond-level protocol verification, and the simulator is used only for one-shot grading rather than interactive agent debugging.
\emph{Third}, all existing benchmarks rely on humans to hand-craft circuits, reference firmwares, and test scripts task by task, which cannot scale across dozens of complex datasheets and heterogeneous MCU hardware and frameworks.

To give coding agents the ability to interactively test and verify hardware just like a human engineer, we present {\labname}, an interactive virtual hardware environment, shown in Fig.~\ref{fig:overview}.
{\labname} emulates the target MCU at the instruction level, so it executes the exact same compiled binary that would be flashed onto a physical board, and connects the processor through a unified peripheral interface to behavioral peripheral models, an analog circuit simulator, and virtual WiFi and BLE network endpoints.
All hardware interaction goes through a standardized, timestamped syntax (\texttt{set}, \texttt{do}, \texttt{read}, and \texttt{assert}) that is bound to the CPU emulator's internal clock, with the test time origin aligned to the entry of the user application function across boards.
With {\labname}, any coding agent can compile, flash, and interactively test its binary, perturb sensor values, press buttons, and inspect component states at millisecond resolution, all without access to physical hardware.

To systematically evaluate both {\labname} and the agents that use it, we also build {\benchname}, a cross-platform benchmark on top of {\labname} that reuses the same syntax to grade a submitted binary against hidden, timed test vectors checking hardware state rather than compilation success or self-printed logs.
Building and verifying peripheral models and tasks by hand across dozens of datasheets and heterogeneous MCU-framework pairs would be a severe engineering bottleneck, so we design a five-stage multi-agent workflow that synthesizes peripheral models and cross-platform tasks from device datasheets and open-source driver libraries.
The workflow places lightweight human checkpoints at the start to approve the hardware specification and at the end to verify reference firmwares on physical boards.
Between these checkpoints, context-isolated builder, tester, and cross-model reviewer agents implement and test each peripheral model against real third-party driver libraries, while a blind trial-solve loop spawns independent solver agents to expose and automatically repair defects in task prompts, test vectors, peripheral models, and the underlying CPU emulators.
Using this workflow, we build 41 peripheral models and 598 tasks across 6 MCU-framework pairs, with 1,485 input stimulus actions and 2,749 hardware assertions in total.

To measure what {\labname} contributes to embedded development, we build a containerized coding agent and expose {\labname} to it at three cumulative levels:
(\emph{i}) no {\labname} ({\myLevelOne}, compiler feedback only),
(\emph{ii}) passive {\labname} ({\myLevelTwo}, serial logs from execution), and
(\emph{iii}) full {\labname} ({\myLevelThree}, active stimulus injection and state inspection).
Running 8 closed- and open-weight models over all 598 tasks at each level yields 14,352 graded trajectories.
Our results show that although agents compile 99.7\% of tasks under {\myLevelOne}, 19.9\% of those compiled firmwares fail hardware assertions at runtime (up to 73\% on individual targets).
Passive serial logs in {\myLevelTwo} yield a net pass rate gain of only 0.9\%, as static idle outputs mislead agents into breaking 6.1\% of previously passing tasks while fixing 7.1\%.
In contrast, full {\labname} in {\myLevelThree} raises the overall pass rate to 89.1\% ($+$4.9\% to $+$14.7\% per model, and up to $+$31\% on individual targets), with the largest gains on the lower-level ESP-IDF ($+$11.6\%) and Zephyr RTOS ($+$14.5\%) targets, and its grading agrees with physical hardware-in-the-loop grading on 98.3\% of sampled firmwares.
The same generic agent with full {\labname} also outperforms prior embedded agent frameworks on identical tasks and models, reaching 90--99\% average pass rate versus 77--90\% for EmbedAgent's compiler-repair loop~\cite{xu2026embedagent} and 67--82\% for AutoEmbed~\cite{yang2026autoembed}; moreover, the compile-only and serial self-check criteria used by prior work falsely accept 25--39\% of broken firmwares.
Interactive hardware access further lets lightweight models catch up with frontier ones: GPT-6~Luna with {\myLevelThree} matches the {\myLevelOne} accuracy of Grok~4.6 and Gemini~3.8~Flash at 30--70$\times$ lower cost.

\begin{table*}[!t]
\centering
\vspace{-3.0mm}
\caption{Comparison of existing embedded coding agents with their evaluation platforms and {\labnamebf}.}
\label{tab:benchmark_comp}
\vspace{-3.0mm}
\footnotesize
\setlength{\tabcolsep}{5pt}
\renewcommand{\arraystretch}{1.15}
\newcommand{\tcell}[1]{\begin{tabular}[c]{@{}c@{}}#1\end{tabular}}
\newcommand{\lcell}[1]{\begin{tabular}[c]{@{}l@{}}#1\end{tabular}}
\resizebox{\linewidth}{!}{%
\begin{tabular}{lcccccccc}
\toprule
 & \multicolumn{5}{c}{\textbf{Hardware \& Simulation Scope}} & \multicolumn{3}{c}{\textbf{Evaluation \& Scaling}} \\
\cmidrule(lr){2-6} \cmidrule(lr){7-9}
\lcell{\textbf{Platform /}\\\textbf{Benchmark}} & \textbf{MCU/Boards} & \textbf{Frameworks} & \textbf{Peripherals} & \textbf{Execution} & \tcell{\textbf{Timing}\\\textbf{Res.}} & \tcell{\textbf{Evaluation}\\\textbf{Oracle}} & \tcell{\textbf{Agent}\\\textbf{Feedback}} & \textbf{Scaling} \\
\midrule
IoTPilot~\cite{gong2025programming} & -- & \tcell{RIOT, Zephyr,\\Contiki, FreeRTOS} & \tcell{3 (I2C, GPIO,\\WiFi/UDP)} & Compile only & -- & \tcell{Compile +\\static API} & \tcell{Compile\\only} & Manual \\
\addlinespace[3pt]
AutoEmbed~\cite{yang2026autoembed} & \tcell{ATmega328P, Nucleo-L4,\\RP2040, ESP32} & Arduino & \tcell{71 (I2C, SPI, UART,\\1-Wire, Analog, GPIO)} & Physical only & -- & \tcell{Static API +\\self-print log} & \tcell{Passive serial\\logging} & Manual \\
\addlinespace[3pt]
EmbedAgent~\cite{xu2026embedagent} & \tcell{ATmega328P, ESP32,\\RPi Pico} & \tcell{Arduino, ESP-IDF,\\MicroPython} & 9 (GPIO only) & \tcell{Closed-source\\sim (Wokwi)} & 2,000\,ms & \tcell{GPIO\\pin state} & \tcell{1-Shot /\\Build-Fix} & Manual \\
\addlinespace[3pt]
IoT-SkillsBench~\cite{li2026skilled} & \tcell{ATmega2560, ESP32-S3,\\nRF52840} & \tcell{Arduino, ESP-IDF,\\Zephyr} & \tcell{23 (I2C, SPI, UART,\\1-Wire, Analog, GPIO)} & Physical only & -- & \tcell{Physical\\HIL check} & \tcell{1-Shot\\(no feedback)} & Manual \\
\addlinespace[3pt]
\lcell{ {\labnamebf} \\ \textbf{(this work)}} & \tcell{\textbf{ATmega328P, ATmega2560,}\\\textbf{ESP32-S3,}\\\textbf{nRF52840}} & \tcell{\textbf{Arduino,}\\\textbf{ESP-IDF,}\\\textbf{Zephyr}} & \tcell{\textbf{41 (I2C, SPI, UART,}\\\textbf{1-Wire, Analog, GPIO,}\\\textbf{WiFi, BLE)}} & \tcell{\textbf{Deterministic}\\\textbf{sim + SPICE}\\\textbf{+ physical}} & \textbf{$<$0.1\,ms} & \tcell{\textbf{Timed}\\\textbf{hardware}\\\textbf{assertions}} & \tcell{\textbf{Active}\\\textbf{stimulus}} & \tcell{\textbf{Self-}\\\textbf{evolving}} \\
\bottomrule
\end{tabular}}
\vspace{-2.0mm}
\end{table*}

In summary, we make the following contributions in this paper:
\begin{itemize}[leftmargin=*, topsep=2pt, itemsep=1pt]
    \item
    \textbf{An interactive virtual hardware environment ({\labnamebf})}:
    We design and build a multi-architecture virtual hardware environment that couples instruction-level CPU emulation (\mycode{avr8js}, \mycode{qemu}, and \mycode{renode}) with 41 peripheral models, \mycode{ngspice} analog circuit solving, and virtual WiFi/BLE network peers under a single deterministic virtual clock.
    A timestamped \mycode{set}/\mycode{do}/\mycode{read} command-line interface exposes this hardware stack to any coding agent, so the agent can flash its compiled binary, perturb sensor inputs, press buttons, and inspect component states at millisecond resolution without access to a physical board.
    \item
    \textbf{A cross-platform benchmark with a scalable construction workflow ({\benchnamebf})}:
    On top of {\labname}, we build a benchmark of 598 tasks over 41 peripherals across 6 MCU-framework pairs, each graded by hidden timed test vectors that assert on hardware state rather than on compilation or self-printed logs.
    To scale the benchmark without manual authoring, we introduce an automated multi-agent workflow that pairs datasheets with community-verified driver libraries and uses context-isolated testing and blind trial solving to synthesize peripheral models and cross-platform tasks with only 4.6\,min of human effort per task.
    \item
    \textbf{An empirical study of embedded agent debugging}:
    We build a containerized coding agent and expose {\labname} to it at three levels (none, passive, and full).
    Across 8 models, 598 tasks, and the three levels (14,352 graded trajectories), we characterize how coding agents use hardware feedback, why passive serial logs fail on input-driven peripherals, and how active stimulus injection unlocks multi-turn hardware debugging, raising the overall pass rate to 89.1\% (up to $+$31\% on individual targets).
    We further show that a generic agent with full {\labname} outperforms prior embedded agent frameworks on identical tasks and models, and that interactive hardware access allows lightight models to match frontier models running without it at 30--70$\times$ lower cost.
\end{itemize}
\noindent
\emph{To the best of our knowledge, {\labname} is the first closed-loop virtual hardware environment that allows coding agents to develop, test, and debug embedded firmware across heterogeneous MCUs, analog circuits, and wireless networks without physical hardware, and the first to be evaluated with a cross-platform benchmark that scales to new peripherals and targets with minutes of human effort per task.}
{\labname}, {\benchname}, the scaling workflow, and all trajectories are open-sourced at \url{https://embedgym.github.io}.

\section{Related Work and Motivation}
\label{sec:related}
\label{sec:motivation}

\myparatight{Coding agents and benchmarks.}
Code generation benchmarks have progressed from function-level synthesis~\cite{chen2021evaluating,austin2021program,hendrycks2021measuring,lai2023ds,liu2023your} to repository- and terminal-level benchmarks~\cite{jimenez2024swe,jain2025livecodebench,zhuo2025bigcodebench,merrill2026terminal,frontiercode2026,cursorbench2026}, alongside multi-turn agent frameworks such as SWE-agent~\cite{yang2024sweagent,wang2025openhands,zhang2024autocoderover,xia2025demystifying} and \mycode{mini-swe-agent}~\cite{minisweagent}. In these general software settings, agents execute code and unit tests directly on the host machine to diagnose and repair bugs in a closed loop~\cite{fan2023automated,shinn2023reflexion,chen2024teaching,xia2024automated}. Embedded firmware, by contrast, is cross-compiled for specific MCUs and interacts with the physical world through hardware registers, analog circuits, and communication buses. {\labname} and {\benchname} bring closed-loop interactive testing to embedded systems by virtualizing both the MCU and peripherals.

\myparatight{LLMs for embedded and IoT systems.}
Recent work has investigated LLM-based code generation and benchmarking for embedded and IoT applications~\cite{xu2024penetrative,ouyang2024llmsense,gong2025programming,yang2026autoembed,xu2026embedagent,li2026skilled}, but existing platforms fall short across hardware scope, evaluation oracles, and scalability (Table~\ref{tab:benchmark_comp}). IoTPilot~\cite{gong2025programming} grades IoT programs using compilation status and static matching of reference API names in the source text, which rejects valid alternative implementations while overlooking runtime hardware bugs. AutoEmbed~\cite{yang2026autoembed} flashes compiled code onto physical boards, but evaluates correctness by checking LLM-inserted serial print statements while sensors remain at static ambient readings. IoT-Skills\-Bench~\cite{li2026skilled} evaluates one-shot firmware generation (with domain skill prompts) on physical boards, providing no runtime hardware feedback to the agent during development. EmbedAgent~\cite{xu2026embedagent} drives the browser-based Wokwi~\cite{wokwi} simulator externally across nine basic GPIO components with a mandatory 2\,s hold delay per state change, and evaluates only single-turn (\mycode{1-Shot}) or compiler-repair (\mycode{Build-Fix}) generation without interactive simulator access.
All prior benchmarks also rely on humans to hand-craft circuits, reference firmwares, and test scripts.

\myparatight{MCU emulation and virtual prototyping.}
Instruction-level emulators such as QEMU~\cite{bellard2005qemu,lcgamboa_qemu}, Renode~\cite{renode}, and \mycode{avr8js}~\cite{avr8js}, as well as firmware re-hosting frameworks~\cite{zaddach2014avatar,gustafson2019toward,clements2020halucinator,feng2020p2im,zhou2021automatic,mera2021dice,scharnowski2022fuzzware}, execute target binaries for individual processor families, while web-based simulators such as Wokwi~\cite{wokwi} and Velxio~\cite{velxio} offer visual circuit canvases for interactive prototyping. However, standalone CPU emulators lack external sensor models and analog circuit solvers, and browser-based prototyping tools are designed for manual GUI interaction rather than deterministic, cross-platform agent evaluation. {\labname} unifies multi-architecture CPU emulation with shared peripheral state machines, \mycode{ngspice} analog circuit simulation, and virtual WiFi and BLE peers under a deterministic virtual clock with aligned application-entry timing.

\begin{figure}[!t]
\centering
\includegraphics[width=\linewidth]{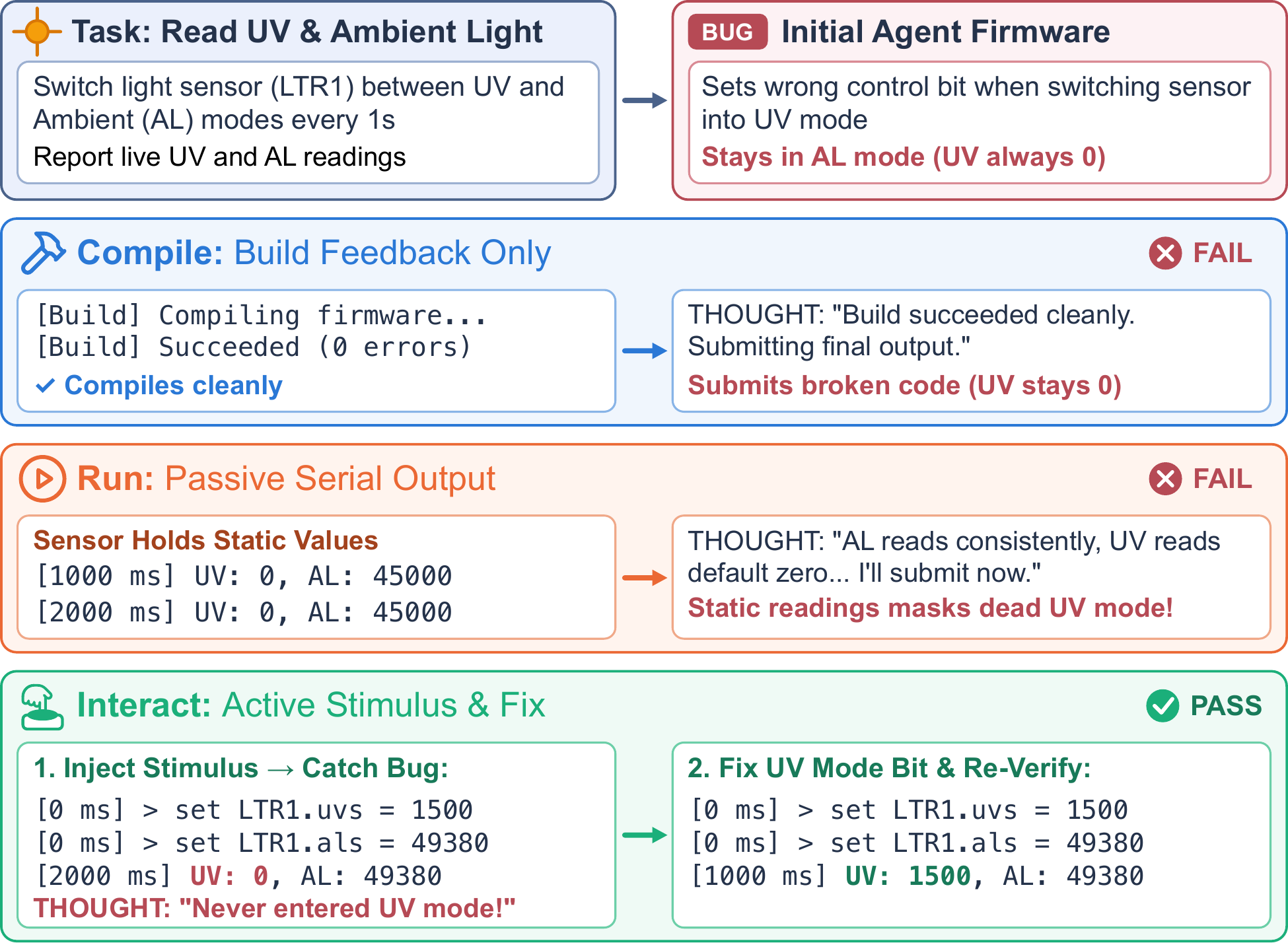}
\vspace{-3.0mm}
\caption{Agentic workflow on a UV and ambient light sensor task.
Faulty firmware passes the \texttt{Compile} and \texttt{Run} stages but is only caught, and then fixed, in the \texttt{Interact} stage.}
\label{fig:motivation_trace}
\vspace{-5.0mm}
\end{figure}

\begin{figure*}[!t]
\centering
\includegraphics[width=.88\linewidth]{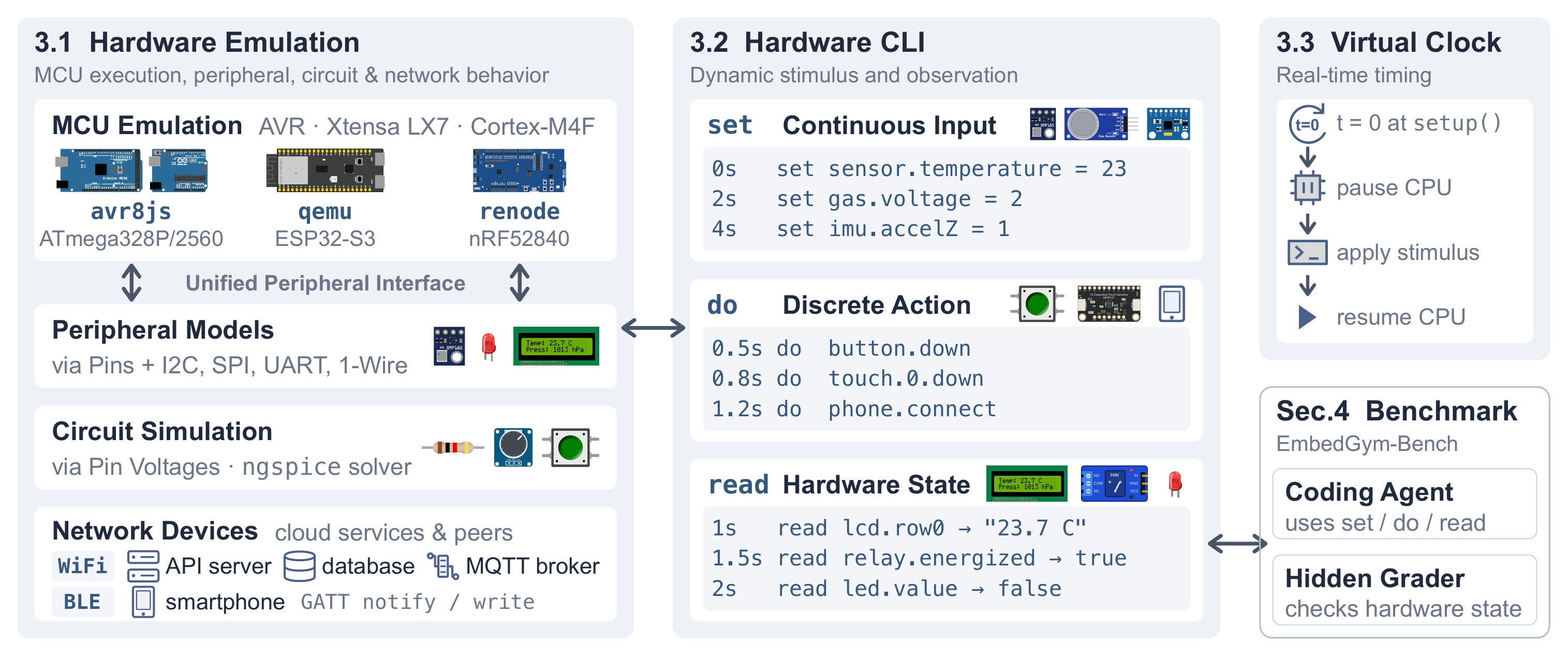}
\vspace{-2.0mm}
\caption{{\labnamebf}'s interactive virtual hardware environment: Instruction-level MCU emulators connected through a unified peripheral interface to 41 peripheral models, an \mycode{ngspice} circuit simulator, and virtual network devices; a text-based hardware interaction CLI (\mycode{set}, \mycode{do}, and \mycode{read}); and deterministic virtual-clock stepping with application-entry time alignment.}
\label{fig:system_arch}
\vspace{-3.0mm}
\end{figure*}

\myparatight{A motivating example.}
Fig.~\ref{fig:motivation_trace} illustrates why the compile-only paradigm (used by IoTPilot~\cite{gong2025programming}) and the passive serial execution paradigm (used by AutoEmbed~\cite{yang2026autoembed}) fail on a real embedded task. The task requires the firmware to switch an optical sensor (\texttt{LTR1}) between ultraviolet (UV) and ambient light (AL) modes every second and report live UV and AL readings. In its initial firmware, the agent writes the wrong control bit when switching into UV mode, leaving the sensor stuck in AL mode so that the UV reading is always zero.
Under compiler feedback only (\texttt{Compile}), the firmware builds with zero errors and the agent immediately submits the broken code. When the compiled binary is executed to collect passive serial output (\texttt{Run}), the console repeatedly prints \texttt{UV: 0, AL: 45000}.
Because a sensor sitting in a static indoor environment naturally reads zero UV without an external UV source, this stationary reading masks the broken UV mode and convinces the agent that its code works properly.
In contrast, under active hardware interaction (\texttt{Interact}), the agent perturbs the sensor with non-zero UV and ambient light inputs (\texttt{set LTR1.uvs = 1500} and \texttt{set LTR1.als = 49380}) at 0\,\text{ms}.
Observing that the console still prints \texttt{UV: 0, AL: 49380} at 2,000\,{ms}, the agent diagnoses that the sensor never entered UV mode, fixes the control register bit, and re-verifies \texttt{UV: 1500, AL: 49380} before submitting.

\section{{\labnamebf}: An Interactive Virtual Hardware Environment for Embedded Coding Agents} 
\label{sec:system}

When a human engineer develops embedded firmware, getting the code to compile is only the first step. Before the firmware can be trusted on real hardware, the engineer verifies it against four progressively stricter requirements:
\emph{Compilation and MCU execution (\textbf{R1})}: the compiled binary boots and executes machine instructions on the target microcontroller without crashing;
\emph{Peripheral, circuit, and network behavior (\textbf{R2})}: the running binary configures hardware registers correctly and communicates properly with external sensors, analog circuits, and wireless endpoints;
\emph{Dynamic stimulus-response logic (\textbf{R3})}: the firmware responds correctly when sensor values change, buttons are pressed, or knobs are turned, so that bugs hidden at static idle defaults are exposed; and
\emph{Real-time timing accuracy (\textbf{R4})}: input sampling, bus protocols, and output responses meet sub-millisecond timing requirements.
A human engineer checks \textbf{R2}--\textbf{R4} at a benchtop with the board, a wired circuit, and a few instruments.
A coding agent, in contrast, works only through a text terminal and cannot touch or see physical sensors or components.
{\labname} provides this capability without physical hardware: it emulates the MCU and its surrounding hardware components and exposes all four requirements through a CLI.

As shown in Fig.~\ref{fig:system_arch}, {\labname} consists of three layers.
The \emph{hardware emulation layer} (\S\ref{sec:sys_sim}) executes the agent's compiled binary on an instruction-level CPU emulator (\mycode{avr8js}, \mycode{qemu}, or \mycode{renode}, depending on the target MCU) and connects the emulated processor, through a unified peripheral interface, to 41 peripheral models written once in Python, an \mycode{ngspice} analog circuit solver, and virtual WiFi and BLE endpoints. This layer covers \textbf{R1} and \textbf{R2}.
The \emph{hardware CLI} (\S\ref{sec:sys_instrumentation}) wraps this hardware stack behind three timestamped commands (\mycode{set}, \mycode{do}, and \mycode{read}), so that any coding agent can compile, flash, and interactively test its binary in a closed loop, covering \textbf{R3}.
The \emph{virtual clock} (\S\ref{sec:sys_timing}) binds every CLI action to the CPU emulator's internal clock and aligns the time origin to the entry of the user application function across platforms, covering \textbf{R4}.
The coding agent and {\benchname} sit outside {\labname} and interact with it only through the CLI (see Fig.~\ref{fig:overview}): during development the agent issues \mycode{set}, \mycode{do}, and \mycode{read} commands of its own choosing, and during grading {\benchname} replays a hidden test vector through the same interface. \S\ref{sec:sys_impl} describes the implementation across 6 MCU-framework pairs.

\subsection{Hardware Emulation: MCU, Peripherals, Circuits, and Networks}
\label{sec:sys_sim}
 
\myparatight{MCU emulation.}
To cover \textbf{R1} across representative embedded architectures, {\labname} builds on three instruction-level CPU emulators: \mycode{avr8js}~\cite{avr8js} for 8-bit AVR processors (e.g., on ATmega328P and ATmega2560), \mycode{qemu}~\cite{bellard2005qemu,lcgamboa_qemu} for 32-bit Xtensa LX7 processors (e.g., on ESP32-S3), and \mycode{renode}~\cite{renode} for 32-bit Arm Cortex-M4F processors (e.g., on nRF52840).
Because each emulator executes the target architecture's instruction set and memory-mapped registers directly, the agent compiles its code with the standard toolchain (\mycode{arduino-cli}, \mycode{esp-idf}, or \mycode{zephyr}) and runs the exact binary that would be flashed onto a physical board.
 
\myparatight{Peripheral models.}
CPU emulators verify instruction execution, but each one is written in a different language and models only the processor chip itself.
If sensors, actuators, and displays were integrated directly into a specific CPU emulator, supporting a new MCU would require rewriting every peripheral model.
We therefore decouple CPU emulation from peripheral modeling through a \emph{unified peripheral interface}.
On the processor side, each CPU emulator translates GPIO transitions, timer events, and bus transfers (I2C, SPI, UART, and 1-Wire) into standardized events (\emph{Pins + Bus Data}).
On the peripheral side, each model implements the component's internal registers, command state machine, and conversion formulas once from its datasheet, and connects to any of the three emulators through this interface.
For example, when the firmware queries a temperature sensor over I2C, the peripheral model checks its configuration registers, updates its internal state, and returns the resulting bytes exactly as the physical chip would.
 
\myparatight{Circuit simulation.}
Beyond digital bus protocols, passive and analog components (e.g., resistors, knobs, and buttons) exhibit circuit-level behaviors expressed as continuous pin voltages, which depend on how the entire circuit is wired.
To model these behaviors, {\labname} connects the MCU and peripheral pins through the unified interface (\emph{Pin Voltages}) to an \mycode{ngspice} circuit solver. 
Whenever a pin state, knob position, or button state changes, the solver recomputes the node voltages across the circuit netlist and feeds them back to the MCU's analog and digital input pins.
This exposes circuit-level bugs such as floating inputs, inverted pin polarities, and wrong voltage-divider scaling that no bus-level model can catch.
 
\myparatight{Network devices.}
Finally, for MCUs that communicate off-board over WiFi or BLE, {\labname} connects the emulated radio through the unified interface (\emph{WiFi/BLE}) to virtual network devices and endpoints, including a virtual smartphone, an API server, a database, and an MQTT broker.
Through these endpoints, the firmware can establish wireless connections, exchange packets, and interact with cloud services end-to-end inside the environment. Together, the peripheral models, circuit solver, and network endpoints cover \textbf{R2}.

\subsection{Hardware Interaction via CLI}
\label{sec:sys_instrumentation}
 
With the MCU and its external hardware running inside {\labname}, \textbf{R3} calls for actively driving physical inputs and observing how the hardware responds, just as an engineer does at a benchtop. However, sensors, buttons, and displays have diverse physical behaviors, and an agent operating in a text terminal cannot manipulate them without a standardized interface.
We abstract benchtop interactions into three timestamped operations:
\mycode{set} changes a continuous physical property such as temperature or light intensity;
\mycode{do} triggers a discrete action such as pressing a button or sending a packet from a network peer; and
\mycode{read} returns hardware states such as pin levels, bus traffic, display text, or serial output.
During development, the agent compiles its firmware and runs it through a CLI tool together with a sequence of \mycode{set}, \mycode{do}, and \mycode{read} actions of its own choosing, for example \mycode{set LTR1.uvs = 1500} at 0\,ms followed by \mycode{read} at 1,000\,ms (Fig.~\ref{fig:motivation_trace}).
This lets the agent test input-dependent logic, diagnose runtime bugs, and revise its code in a closed loop.
 
The interface is deliberately agent-agnostic: it requires only the ability to run a shell command and read its output, so any coding agent that can do so can use {\labname} without modification.
The same property lets {\benchname} (described in \S\ref{sec:workflow}) reuse the exact same syntax to grade the agent's final firmware, pairing \mycode{set} and \mycode{do} stimuli with \mycode{assert} checks on expected hardware states and serial outputs.
It also means that a physical-board rig exposing the same three commands could stand in for {\labname} transparently, for both the agent and the benchmark.

\subsection{Timing Control and Alignment} \label{sec:sys_timing}

\begin{figure}[!t]
\centering
\includegraphics[width=0.70\linewidth]{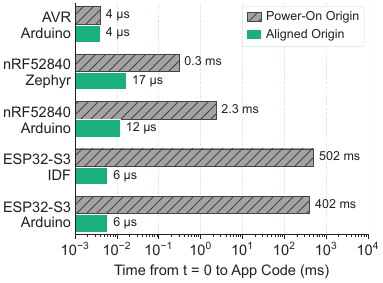}
\vspace{-3.0mm}
\caption{Time from $t = 0$ to the first application function for each MCU and framework (log scale). Gray bars show the sum of ROM boot, bootloader, and runtime startup time; green bars show the aligned time origin.}
\label{fig:timing_contract}
\vspace{-3.0mm}
\end{figure}

When an agent or test vector schedules \mycode{set}, \mycode{do}, \mycode{read}, or \mycode{assert} actions at specific timestamps, {\labname} must execute those timed events accurately and consistently across MCUs. 
Prior work such as EmbedAgent~\cite{xu2026embedagent} drives an external simulator (Wokwi) using the host machine's wall-clock time.
To mask host CPU load and network jitter, it must hold each state for 2 seconds, which prevents testing millisecond-level protocols and sampling intervals.
{\labname} avoids wall-clock jitter by binding the test runner directly to the CPU emulator's internal clock: whenever the emulated clock reaches an action's timestamp, the runner pauses the CPU, applies the \mycode{set} or \mycode{do} input or evaluates the \mycode{read} or \mycode{assert} check, and then resumes execution.
This allows reproducible, millisecond-level testing regardless of host machine speed.

By default, an emulator's time origin ($t = 0$) is at hardware power-on or reset.
However, we observe that the startup time from power-on to the entry of the user's application function (\mycode{setup()}, \mycode{main()}, or \mycode{app\_main()}) spans five orders of magnitude across platforms, as shown in Fig.~\ref{fig:timing_contract}: AVR takes only 4\,$\mu$s and nRF52840 takes {0.3--2.3}\,{ms}, whereas ESP32-S3 takes {402--502}\,ms to finish ROM boot, bootloader execution, and runtime startup.
If test timestamps were measured from power-on, the same interaction sequence or test vector would trigger at completely different stages of execution across boards, and would fire on ESP32-S3 before the user's code even begins.
We resolve this mismatch at its source by aligning the test time origin $t = 0$ to the entry of the user application function rather than power-on.
We apply a lightweight runtime wrapper that emits a synchronization signal right before calling the user application function.
When the emulator detects this signal, it resets the test clock to $t = 0$ and clears prior boot logs.
As shown in Fig.~\ref{fig:timing_contract}, this reduces the cross-platform starting offset from as high as 100s of milliseconds to less than 20 microseconds.

\subsection{Implementation of {\labnamebf}}
\label{sec:sys_impl}

\begin{table}[!t]
\centering
\caption{Implementation breakdown in programming languages and lines of code (LoC) of {\labnamebf}.}
\label{tab:impl_loc}
\vspace{-2.0mm}
\small
\setlength{\tabcolsep}{4pt}
\begin{tabular*}{\linewidth}{@{\extracolsep{\fill}}llr@{}}
\toprule
\textbf{Component} & \textbf{Language(s)} & \textbf{LoC} \\
\midrule
\mycode{avr8js} (AVR) wrapper \& bridge & TS & 1.2K \\
\mycode{qemu} (ESP32-S3) patch \& bridge & C, Python & 2.1K \\
\mycode{renode} (nRF52840) patch \& bridge & C\#, Python, TS & 2.2K \\
Shared MCU worker bridge & TS, Python & 1.4K \\
\midrule
41 peripheral models & Python, TS & 11.6K \\
Peripheral datasheet verification suite & Python & 6.8K \\
Pin/bus core \& \mycode{ngspice} circuit bridge & Python, TS & 2.3K \\
WiFi \& BLE network endpoints & Python, C, TS & 1.9K \\
\midrule
Hardware CLI, clock runner \& $t=0$ hooks & TS, Shell, Py, C & 2.6K \\
\midrule
\textbf{Total ({\labnamebf})} & \textbf{Py / TS / C / C\#} & \textbf{32.3K} \\
\bottomrule
\end{tabular*}
\vspace{-2.0mm}
\end{table}

We implement {\labname} in approximately 32.3K lines of Python, TypeScript, C, C\#, and Shell code, as summarized in Table~\ref{tab:impl_loc}.
Across the 6 MCU-framework pairs introduced in \S\ref{sec:sys_sim}, we run ATmega328P and ATmega2560 (Arduino) on unmodified \mycode{avr8js}~\cite{avr8js} via a TypeScript wrapper; ESP32-S3 (Arduino and ESP-IDF) on a customized \mycode{qemu} fork derived from~\cite{lcgamboa_qemu,velxio}; and nRF52840 (Arduino and Zephyr RTOS) on \mycode{renode}~\cite{renode}.
Because these emulators are written in different languages and model only the corresponding processor chip(s), we instrument \mycode{qemu} (compiled as a shared C library) and \mycode{renode} (via C\# hooks and its Python runtime interface) with custom callbacks on their GPIO, ADC, PWM, and bus controllers (I2C, SPI, and UART).
To share a single peripheral implementation across all three emulators, \mycode{qemu} and \mycode{renode} invoke our Python peripheral models directly in the backend, while the TypeScript-based \mycode{avr8js} executes the same Python models through the WebAssembly Python runtime (\mycode{Pyodide}), alongside a WebAssembly build of the \mycode{ngspice} circuit solver.
For ESP32-S3's WiFi capability, we connect \mycode{qemu}'s virtual network interface to a Docker container running MQTT, HTTP/WebSocket, CoAP, NTP, and PostgreSQL services; for nRF52840's BLE capability, we pair \mycode{renode}'s emulated radio with a virtual smartphone running a second emulated nRF52840 Zephyr HCI controller and the Python \mycode{Bumble} BLE host in synchronized virtual time.
The time-alignment hooks (\S\ref{sec:sys_timing}) are pre-configured via linker wrappers in the build toolchains, so compiled binaries link them automatically without changes to user code.

\section{Design of {\benchnamebf}}
\label{sec:workflow}

\begin{figure}
\centering
\includegraphics[width=\linewidth]{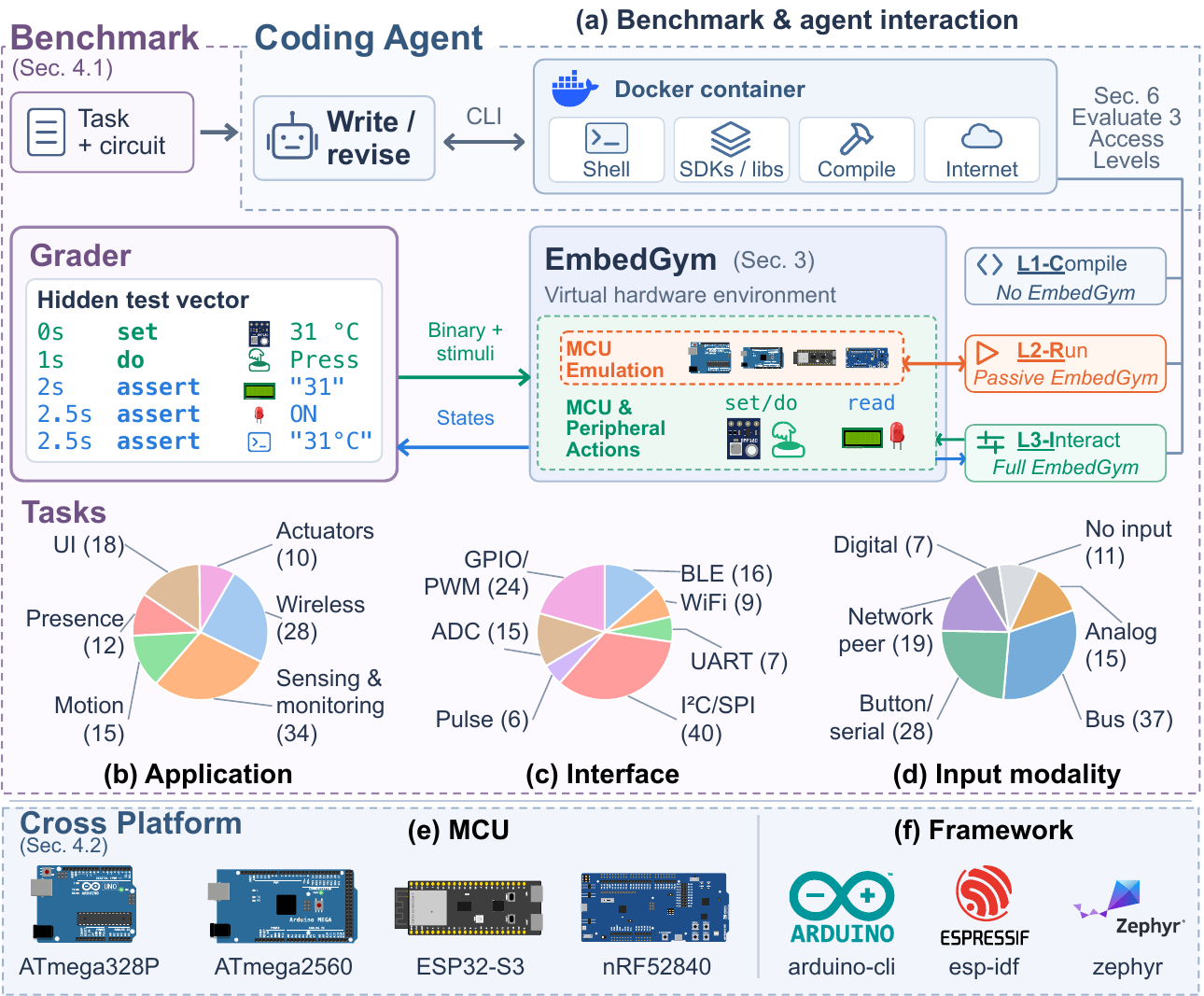}
\caption{Architecture and composition of {\benchnamebf}: (a)~benchmark and three levels of agent interaction; distribution of the tasks by (b)~application domain, (c)~hardware interface, and (d)~input modality; and cross-platform support across (e)~four MCUs and (f)~three frameworks.}
\label{fig:bench_arch}
\end{figure}

To systematically measure how much {\labname} helps coding agents and to evaluate those agents on realistic embedded firmware development, we build {\benchname}, as shown in Fig.~\ref{fig:bench_arch}.
Designing the benchmark requires solving two challenges: how an individual task gives the agent realistic hardware context while grading its compiled binary against ground-truth hardware states without leaking the test cases (\S\ref{sec:bench_task}); and how the benchmark compares agents fairly across heterogeneous MCU architectures and software frameworks that differ in pinouts, voltage rails, and driver APIs (\S\ref{sec:bench_suite}). A third challenge, constructing and extending the suite at scale, is addressed by the multi-agent workflow in \S\ref{sec:impl}.

\subsection{Task Structure and Hardware-State Grading} \label{sec:bench_task}

Each task in {\benchname} separates the agent-facing development workspace from a hidden grading harness (Fig.~\ref{fig:bench_arch}(a)).
On the agent side, the workspace provides a natural-language specification of the required functional behavior, sampling intervals, and output formats, along with a target-specific pin wiring table (or a devicetree overlay on Zephyr RTOS), vendor build toolchains, and {\labname}'s CLI tools (\texttt{set}, \texttt{do}, and \texttt{read}).
The agent iteratively writes code, compiles binaries, and runs interactive tests in {\labname} until it submits a final compiled binary.
Throughout development, the full circuit netlist, reference firmware, and grading test vectors remain isolated outside the agent's container so the agent cannot inspect or hardcode the test cases.

To grade the submitted binary, the harness loads the firmware into a clean {\labname} instance wired with the task's hidden circuit netlist and executes a predefined test vector using the time-aligned \texttt{set}, \texttt{do}, and \texttt{assert} operations (see \S\ref{sec:sys_instrumentation} and \S\ref{sec:sys_timing}).
For example, a test vector can \texttt{set} a temperature sensor to $31\,^\circ$C at $t = 0\,\text{s}$, \texttt{do} a button press at $t = 1\,\text{s}$, and \texttt{assert} at $t = 2\,\text{s}$ and $2.5\,\text{s}$ that an LCD display shows \texttt{"31"}, an alarm LED pin is driven high, and the serial port prints \texttt{"31 °C"}.
For wireless tasks, assertions also query the virtual cloud database, MQTT broker, HTTP server, or BLE peer to verify end-to-end data exchange. A task is graded as passed if and only if the binary compiles cleanly, boots without crashing, and satisfies every assertion across all timestamps in the test vector.

\subsection{Cross-Platform Task Design and Suite} \label{sec:bench_suite}

A central design goal of {\benchname} is to allow for evaluating agents across diverse MCU architectures (8-bit AVR, 32-bit Xtensa LX7, and 32-bit Arm Cortex-M4F) and software frameworks (Arduino, ESP-IDF, and Zephyr RTOS) on equal footing (Fig.~\ref{fig:bench_arch}(e)--(f)).
On real hardware, porting a task across boards requires changing pin numbers, voltage levels, ADC resolutions, register APIs, and RTOS device bindings.
To evaluate these different implementations against a single ground truth, we decouple \emph{target-specific hardware wiring} from a \emph{target-agnostic behavioral test vector}. Each target platform has its own circuit netlist and pin mapping table tailored to its board layout, while all targets share a single test vector written in terms of component-level stimuli and assertions (\S\ref{sec:sys_instrumentation}) and synchronized to the user application entry $t = 0$ (\S\ref{sec:sys_timing}).
This design allows the exact same test vector to grade an Arduino sketch on ATmega328P, an ESP-IDF C program on ESP32-S3, and a Zephyr RTOS application on nRF52840.

Following this cross-platform design, {\benchname} comprises 598 tasks across 6 MCU-framework pairs: ATmega328P and ATmega2560 (\mycode{arduino-cli}, $N = 92$ each), ESP32-S3 (\mycode{arduino-cli} and \mycode{esp-idf}, $N = 100$ each), and nRF52840 (\mycode{arduino-cli} and \mycode{zephyr}, $N = 107$ each).
These 598 tasks are instantiated from 117 unique task specifications across 41 peripherals: 91 core peripheral and circuit specifications are shared across all 6 MCU-framework pairs ($91 \times 6 = 546$ tasks), while 26 specifications exercise platform-specific capabilities, including 9 WiFi and cloud networking tasks on the ESP32-S3 with both frameworks ($9 \times 2 = 18$), 16 BLE wireless tasks on nRF52840 with both frameworks ($16 \times 2 = 32$), and one dual-motor line-follower task on the two AVR MCUs ($1 \times 2 = 2$), totaling $546 + 18 + 32 + 2 = 598$ tasks (with 1,485 dynamic input stimuli and 2,749 hardware assertions).
As shown in Fig.~\ref{fig:bench_arch}(b)--(d), the 117 task specifications span three orthogonal dimensions: six application categories (\emph{Sensing \& Monitoring} (34), \emph{Wireless} (28), \emph{UI} (18), \emph{Motion} (15), \emph{Presence} (12), and \emph{Actuators} (10)), seven primary hardware interface categories (\emph{I2C/SPI} (40), \emph{GPIO/PWM} (24), \emph{BLE} (16), \emph{ADC} (15), \emph{WiFi} (9), \emph{UART} (7), and \emph{Pulse} (6)), and six dynamic input modalities (\emph{Bus} (37), \emph{Button/Serial} (28), \emph{Network Peer} (19), \emph{Analog} (15), \emph{No Input} (11), and \emph{Digital} (7)).

\section{Self-Evolving Scaling of {\labbenchnamebf}}
\label{sec:impl}

\begin{figure*}[!t]
\centering
\includegraphics[width=\linewidth]{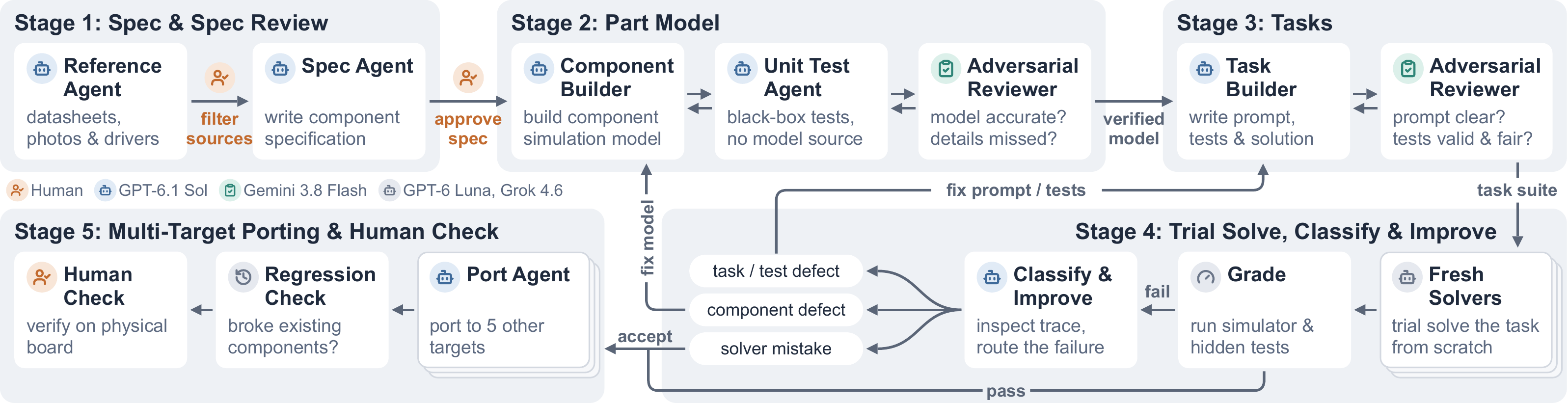}
\vspace{-3.0mm}
\caption{Five-stage self-evolving workflow for constructing {\labnamebf} and {\benchnamebf}: human checkpoints filter reference sources and approve the hardware specification (Stage~1), context-isolated agents build and verify the peripheral model in {\labnamebf} (Stage~2) and synthesize tasks in {\benchnamebf} (Stage~3), a blind trial-solve loop classifies failures to iteratively repair task tests or component models on the first target (Stage~4), and accepted tasks and simulator bindings are ported across the remaining five targets and verified on physical boards (Stage~5).}
\label{fig:workflow}
\vspace{-3.0mm}
\end{figure*}

Scaling our evaluation requires expanding both the peripheral models in {\labname} (\S\ref{sec:system}) and the cross-platform tasks, target circuits, reference firmwares, and calibrated test vectors in {\benchname} (\S\ref{sec:workflow}). Building and verifying both layers manually across dozens of datasheets and six targets would create a severe engineering bottleneck. To automate both peripheral modeling and task creation with minimal human effort, we design a five-stage self-evolving multi-agent workflow (Fig.~\ref{fig:workflow}): (1)~reference collection and specification (\emph{Spec \& Spec Review}) to establish a shared hardware specification for both {\labname} and {\benchname}; (2)~context-isolated component modeling and testing (\emph{Part Model}) to expand {\labname}'s peripheral library; (3)~task synthesis and review (\emph{Tasks}) to construct {\benchname} task suites on the first target; (4)~blind trial solving, failure classification, and improvement (\emph{Trial Solve, Classify \& Improve}) to jointly validate and repair both {\benchname} tasks and {\labname} models; and (5)~multi-target porting and hardware verification (\emph{Multi-Target Porting \& Human Check}) to port and verify both {\labname} bindings and {\benchname} tasks across all remaining platforms.

\myparatight{Stage 1: Reference collection and specification (\emph{Spec \& Spec Review}).}
Because chip datasheets often leave startup sequences or register defaults ambiguous and breakout modules from different vendors vary in pinouts and bus addresses, Stage~1 pairs datasheets with community-verified open-source driver libraries and module photos, and places lightweight human checkpoints before any code is written (Stage~1 in Fig.~\ref{fig:workflow}). Because these driver libraries are already tested on physical hardware, their source code reveals the exact register sequences and timing rules that real firmware relies on, and can later be executed directly against our simulator to verify the peripheral model.
A \emph{Reference Agent} first searches the web to collect candidate module photos, chip datasheets, and driver libraries, and a human screens these candidates to select the correct sources (\emph{Filter Sources}).
The approved files are then passed to the \emph{Spec Agent}, which cross-references the datasheet against the driver code to synthesize a structured hardware specification and flag any ambiguities.
A human reviews and approves this specification (\emph{Approve Spec}) as the ground truth.

\myparatight{Stage 2: Context-isolated component modeling and review (\emph{Part Model}).}
Once the specification is approved, the pipeline builds and verifies the peripheral model in {\labname} autonomously (Stage~2 in Fig.~\ref{fig:workflow}). To prevent an agent from encoding the same misunderstanding of the specification into both the peripheral model and its unit tests, we enforce context isolation and cross-model review. After the \emph{Component Builder} implements the simulation model from the specification, an isolated \emph{Unit Test Agent} writes register-level tests and runs the approved third-party driver libraries unmodified against the component, without access to the model's source code. A \emph{Reviewer} from a different model family then audits both the model and the tests against the specification to catch any missed details or timing bugs before passing the verified component model on to the next stage.

\myparatight{Stage 3: Task synthesis and cross-model review (\emph{Tasks}).}
With the verified peripheral model in place, Stage~3 constructs the corresponding {\benchname} tasks on the first target (Stage~3 in Fig.~\ref{fig:workflow}). The \emph{Task Builder} generates the natural-language task prompt, target circuit netlist, hidden timed test vector (\texttt{set}, \texttt{do}, and \texttt{assert}), and a reference firmware implementation that passes the test vector inside {\labname}. A cross-model \emph{Reviewer} then audits the candidate task suite to verify that the prompt is unambiguous and that the hidden assertions are valid and fair before forwarding the task suite to blind trial solving.

\myparatight{Stage 4: Blind trial solving, classification, and improvement (\emph{Trial Solve, Classify \& Improve}).}
Even after static review, a test vector may still be too loose (passing hardcoded outputs) or too strict (rejecting valid alternative implementations), or may expose subtle edge cases in the peripheral model. Stage~4 runs a closed-loop blind trial-solve process on the first target (Stage~4 in Fig.~\ref{fig:workflow}). Independent \emph{Fresh Solvers} receive only the public task prompt and the {\labname} CLI, with no access to the hidden tests or reference firmware, and their submitted binaries are evaluated by the \emph{Grade} harness. If a solver passes, or once failures are resolved, the task is marked as verified on the first target (\emph{1st Target Verified}). Whenever a solver fails grading, the \emph{Classify \& Improve} agent inspects the execution trace and routes the failure to one of three paths:
\begin{itemize}[leftmargin=*, topsep=2pt, itemsep=1pt]
    \item \textbf{Task / Test Defect:} If a hidden assertion enforces an unstated timing or implementation detail, or if the prompt is ambiguous, the workflow routes back to Stage~3 to repair the prompt or test vector and retry (\emph{Fix Prompt / Tests}).
    \item \textbf{Component Defect:} If valid solver firmware fails due to an inaccurate peripheral model or simulator bug, the workflow routes the defect back to Stage~2 to fix the model or simulator and re-run unit tests (\emph{Fix Model / Sim}).
    \item \textbf{Solver Mistake:} If the solver genuinely wrote incorrect firmware, the benchmark remains unchanged and the task is accepted on the first target (\emph{Accepted Task}).
\end{itemize}

\myparatight{Stage 5: Multi-target porting and hardware check (\emph{Multi-Target Porting \& Human Check}).}
Stabilizing tasks on the first target decouples task design from multi-platform simulator debugging. Once a task's prompt and shared test vector are verified in Stage~4, we hold them fixed and port the peripheral bindings and task circuit to the remaining five targets (\emph{Agent Work} in Stage~5 of Fig.~\ref{fig:workflow}). Because the task assertions are already validated, any failure of a reference firmware on a new target directly exposes missing target bindings or low-level emulator bugs in {\labname}, which the agent patches at their source. To ensure these shared simulator edits do not break existing components, each porting step runs a full regression suite across all affected targets (\emph{Regression Check}). Finally, a human follows a generated wiring sheet to assemble the circuit on a physical board, flash the reference firmware, and confirm its behavior matches the specification (\emph{Human Check}).

\myparatight{Self-evolving pipeline and the construction of {\labbenchnamebf}.}
In the five-stage scaling workflow shown in Fig.~\ref{fig:workflow}, we use GPT-6.1~Sol for reference collection, specification synthesis, model and task building, failure classification, and multi-target porting; Gemini~3.8~Flash for cross-model review in Stages~2 and~3; and GPT-6~Luna and Grok~4.6 as blind trial solvers in Stage~4.
Peripherals and tasks are first synthesized and verified on the ATmega2560 platform before being ported to the remaining five MCU-framework pairs.
Using this workflow, we build 41 peripheral models (\S\ref{sec:system}) and 598 tasks across 6 MCU-framework pairs in {\benchname} (\S\ref{sec:workflow}).

\section{Evaluation}
\label{sec:eval}

\begin{table*}[t]
\centering
\caption{Multi-turn agent pass rates for each target under three cumulative feedback levels: L1-Compile provides build feedback; L2-Run adds simulator execution and serial output; L3-Interact also lets the agent control inputs and inspect component states. For each target and model, levels are compared using only tasks evaluated at all three levels. $N$ denotes the number of tasks per target. The highest pass rates are shown in bold.
({\myLevelOne}: Compile; {\myLevelTwo}: Run; {\myLevelThree}: Interact).}
\label{tab:eval_leaderboard}
\vspace{-2.0mm}
\setlength{\tabcolsep}{2.2pt}
\resizebox{\linewidth}{!}{%
\begin{tabular}{lcccccccccccccccccccccccccccccc}
\toprule
\textbf{MCU} & \multicolumn{5}{c}{\textbf{ATmega328P}} & \multicolumn{5}{c}{\textbf{ATmega2560}} & \multicolumn{10}{c}{\textbf{ESP32-S3}} & \multicolumn{10}{c}{\textbf{nRF52840}} \\
\cmidrule(lr){2-6} \cmidrule(lr){7-11} \cmidrule(lr){12-21} \cmidrule(lr){22-31}
\textbf{Framework} & \multicolumn{5}{c}{\mycode{arduino-cli} ($N = 92$)} & \multicolumn{5}{c}{\mycode{arduino-cli} ($N = 92$)} & \multicolumn{5}{c}{\mycode{arduino-cli} ($N = 100$)} & \multicolumn{5}{c}{\mycode{esp-idf} ($N = 100$)} & \multicolumn{5}{c}{\mycode{arduino-cli} ($N = 107$)} & \multicolumn{5}{c}{\mycode{zephyr} ($N = 107$)} \\
\cmidrule(lr){2-6} \cmidrule(lr){7-11} \cmidrule(lr){12-16} \cmidrule(lr){17-21} \cmidrule(lr){22-26} \cmidrule(lr){27-31}
\textbf{Model} & \textcolor{lvlC}{\myLevelOne} & \textcolor{arrowgray}{$\triangleright$} & \textcolor{lvlR}{\myLevelTwo} & \textcolor{arrowgray}{$\triangleright$} & \textcolor{lvlI}{\textbf{\myLevelThree}} & \textcolor{lvlC}{\myLevelOne} & \textcolor{arrowgray}{$\triangleright$} & \textcolor{lvlR}{\myLevelTwo} & \textcolor{arrowgray}{$\triangleright$} & \textcolor{lvlI}{\textbf{\myLevelThree}} & \textcolor{lvlC}{\myLevelOne} & \textcolor{arrowgray}{$\triangleright$} & \textcolor{lvlR}{\myLevelTwo} & \textcolor{arrowgray}{$\triangleright$} & \textcolor{lvlI}{\textbf{\myLevelThree}} & \textcolor{lvlC}{\myLevelOne} & \textcolor{arrowgray}{$\triangleright$} & \textcolor{lvlR}{\myLevelTwo} & \textcolor{arrowgray}{$\triangleright$} & \textcolor{lvlI}{\textbf{\myLevelThree}} & \textcolor{lvlC}{\myLevelOne} & \textcolor{arrowgray}{$\triangleright$} & \textcolor{lvlR}{\myLevelTwo} & \textcolor{arrowgray}{$\triangleright$} & \textcolor{lvlI}{\textbf{\myLevelThree}} & \textcolor{lvlC}{\myLevelOne} & \textcolor{arrowgray}{$\triangleright$} & \textcolor{lvlR}{\myLevelTwo} & \textcolor{arrowgray}{$\triangleright$} & \textcolor{lvlI}{\textbf{\myLevelThree}} \\
\midrule
\modellogo{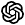}GPT-5.6 Luna & 89\% & & 84\% & & \textbf{92\%} & 85\% & & 84\% & & \textbf{93\%} & 78\% & & 78\% & & \textbf{92\%} & 65\% & & 64\% & & \textbf{78\%} & 63\% & & 63\% & & \textbf{69\%} & 58\% & & 60\% & & \textbf{71\%} \\
\rowcolor{rowband} \modellogo{openai}GPT-5.6 Sol & 92\% & & 93\% & & \textbf{100\%} & 96\% & & 96\% & & \textbf{100\%} & 93\% & & 96\% & & \textbf{99\%} & 90\% & & 89\% & & \textbf{99\%} & 89\% & & 88\% & & \textbf{94\%} & 80\% & & 82\% & & \textbf{93\%} \\
\modellogo{openai}GPT-6 Luna & 89\% & & 90\% & & \textbf{98\%} & 85\% & & 88\% & & \textbf{92\%} & 81\% & & 84\% & & \textbf{94\%} & 76\% & & 85\% & & \textbf{94\%} & 68\% & & 68\% & & \textbf{76\%} & 73\% & & 77\% & & \textbf{82\%} \\
\rowcolor{rowband} \modellogo{openai}GPT-6.1 Sol & 98\% & & 97\% & & \textbf{99\%} & 97\% & & 97\% & & \textbf{100\%} & 93\% & & 94\% & & \textbf{98\%} & 91\% & & 94\% & & \textbf{98\%} & 93\% & & 94\% & & \textbf{99\%} & 90\% & & 90\% & & \textbf{95\%} \\
\modellogo{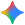}Gemini 3.8 Flash & 93\% & & 96\% & & \textbf{98\%} & 96\% & & 95\% & & \textbf{100\%} & \textbf{96\%} & & 91\% & & \textbf{96\%} & 87\% & & 93\% & & \textbf{99\%} & \textbf{93\%} & & 90\% & & \textbf{93\%} & 79\% & & 79\% & & \textbf{92\%} \\
\rowcolor{rowband} \modellogo{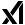}Grok 4.6 & 93\% & & 97\% & & \textbf{99\%} & 92\% & & 92\% & & \textbf{99\%} & 90\% & & 91\% & & \textbf{97\%} & 92\% & & 88\% & & \textbf{95\%} & 85\% & & 86\% & & \textbf{93\%} & 85\% & & 79\% & & \textbf{93\%} \\
\modellogo{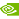}Nemotron 3.5 Super VL & 77\% & & 77\% & & \textbf{84\%} & 77\% & & 78\% & & \textbf{85\%} & 73\% & & 76\% & & \textbf{86\%} & 56\% & & 56\% & & \textbf{68\%} & 59\% & & 64\% & & \textbf{75\%} & 27\% & & 38\% & & \textbf{58\%} \\
\rowcolor{rowband} \modellogo{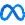}Muse Glimmer 30B & 79\% & & 82\% & & \textbf{96\%} & 85\% & & 80\% & & \textbf{93\%} & 81\% & & 84\% & & \textbf{90\%} & 54\% & & 55\% & & \textbf{73\%} & 68\% & & 64\% & & \textbf{76\%} & 27\% & & 34\% & & \textbf{51\%} \\
\bottomrule
\end{tabular}}
\end{table*}

We evaluate {\labname} on {\benchname} with 598 tasks across 6 MCU-framework pairs, using eight closed- and open-weight models, as shown in Table~\ref{tab:eval_leaderboard}.
We expose {\labname} to the agent at three cumulative hardware feedback levels:
\begin{itemize}[leftmargin=*, topsep=2pt, itemsep=1pt]
\item
\textbf{{\myLevelOne} (Compile, \emph{no} {\labnamebf})}, where the agent has only the toolchain, and build errors are the sole feedback;
\item
\textbf{{\myLevelTwo} (Run, \emph{passive} {\labnamebf})}, where the agent executes the compiled binary on the virtual hardware and reads its serial output, so it can observe the hardware; and
\item
\textbf{{\myLevelThree} (Interact, \emph{full} {\labnamebf})}, where the agent can additionally issue \mycode{set}/\mycode{do} stimuli and \mycode{read} queries, so it can drive inputs and inspect component states while debugging. 
\end{itemize}

\myparatight{Containerized agent environment and test runner.}
To give coding agents full compilation and interactive testing tools in {\labname} while preventing access to the hidden circuit topologies and grading vectors of {\benchname}, we isolate the agent and the simulator in separate Docker containers.
For each task in {\benchname}, the harness spawns a fresh build container for the coding agent, built on \mycode{mini-swe-agent}~\cite{minisweagent}, to ensure a clean environment with no leftover build artifacts, installed libraries, or file traces from previous tasks.
The container includes only the {\benchname} task description, pin wiring table (or Zephyr devicetree overlay), and base toolchain (\mycode{arduino-cli}, \mycode{esp-idf}, or \mycode{zephyr} via \mycode{west}), requiring the agent to search for and install any needed third-party peripheral libraries on demand. 
For cross-platform time alignment (\S\ref{sec:sys_timing}), we pre-install the runtime wrapper in the container's SDK and Arduino core, so that every compiled binary links it automatically without changes to user code.
When the agent invokes the simulation command, the build container sends the compiled binary and the agent's timed \mycode{set}, \mycode{do}, and \mycode{read} actions over a local socket to the {\labname} simulator container, which executes the run and returns only the console logs and requested component states.
Before every interactive run in {\labname} and final grading pass in {\benchname}, the harness resets all peripheral states, cloud databases and brokers, and the virtual BLE peer so earlier debugging actions never pollute later tests.

\myparatight{Baselines.}
On the four Arduino targets in {\benchname}, we compare our agent equipped with {\labname} against two state-of-the-art embedded coding baselines using the same underlying LLMs: EmbedAgent~\cite{xu2026embedagent} and AutoEmbed~\cite{yang2026autoembed} (Fig.~\ref{fig:baseline_comparison}).
For EmbedAgent~\cite{xu2026embedagent}, we evaluate both of its generation modes: \mycode{1-Shot}, which generates firmware in a single turn from the task prompt and pinout without compiler feedback, and \mycode{Build-Fix}, which extends \mycode{1-Shot} with one round of compiler-error repair using build logs.
For AutoEmbed~\cite{yang2026autoembed}, we run its multi-stage pipeline, which iteratively repairs compilation errors and validates the firmware using LLM-inserted serial print statements (\emph{Serial Self-Check}).
We evaluate both baselines on the tasks in {\benchname}, so neither baseline interacts with {\labname}'s stimulus injection or state inspection during development; their final submitted binaries are then graded by the exact same hidden hardware assertions in {\benchname} and verified by a human expert.

\myparatight{Metrics.}
We evaluate performance using three primary metrics:
(1)~\emph{task pass rate (\%)}, the percentage of tasks in {\benchname} where the submitted binary compiles, boots, and satisfies every hidden timed hardware assertion across the test vector;
(2)~\emph{mean API cost per task (\$)} and \emph{interaction steps}, calculated from the input and output token counts of every model call at listed API prices alongside the number of turns spent on codebase exploration, editing, building, and simulation; and
(3)~\emph{false discovery rate (FDR, \%)} and \emph{sim-to-real agreement (\%)}, which measure the percentage of broken firmwares falsely accepted by prior validation criteria and the grading agreement between {\labname} and physical hardware boards (\S\ref{sec:eval_fidelity}).
Each agent run is capped at 100 steps and a \$1.0 token budget, yielding 14,352 graded trajectories across the 8 models, 598 tasks, and three capability levels (4,784 trajectories per level).

\subsection{End-to-End Performance} 
\label{sec:eval_main}

\begin{figure}[!t]
\centering
\includegraphics[width=0.98\linewidth]{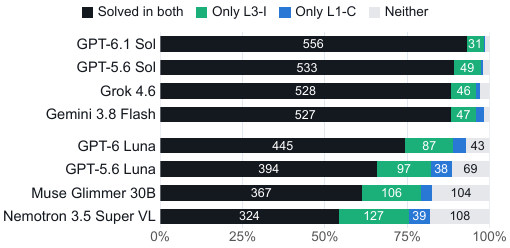}
\vspace{-3.0mm}
\caption{Breakdown of the 598 tasks across all six hardware targets by outcome under {\myLevelOne} and {\myLevelThree} for each model; numbers inside the bars denote task counts.}
\label{fig:solved_by_level}
\vspace{-3.0mm}
\end{figure}

\myparatight{Overall performance across models and targets.}
Table~\ref{tab:eval_leaderboard} and Fig.~\ref{fig:solved_by_level} present the end-to-end pass rates across all eight models and six hardware platforms.
With full interactive access to {\labname} ({\myLevelThree}), the agent achieves an overall pass rate of 89.1\% across all 4,784 runs, with frontier models such as GPT-6.1~Sol and GPT-5.6~Sol reaching 93\%--100\% across targets.
In contrast, under {\myLevelOne}, although the agent successfully builds 99.7\% (4,772/4,784) of the tasks, 19.9\% of them fail hardware assertions at runtime (ranging from 6.7\% to 39.0\% across models and up to 73\% on individual targets). 
Adding passive serial logs in {\myLevelTwo} increases overall pass rate by only 0.9\%, and 13 of the 48 model-platform pairs actually observe lower accuracy than {\myLevelOne}.
Enabling active stimulus and state inspection in {\myLevelThree} consistently outperforms both {\myLevelOne} and {\myLevelTwo} across every model-platform pair ($+$4.9\% to $+$14.7\% per model, and up to $+$31\% on a single target hardware platform).
As shown in Fig.~\ref{fig:solved_by_level}, across the 598 tasks per model, {\myLevelThree} rescues 31--127 tasks that fail under {\myLevelOne} while regressing on only 2--39 tasks, with open-weight models gaining the largest number of newly solved tasks (106 on Muse~Glimmer~30B and 127 on Nemotron~3.5~Super~VL).
The largest performance gain is observed on native SDK and RTOS frameworks.
Specifically, {\myLevelOne} averages 9.2\% lower on ESP-IDF and 12.3\% lower on Zephyr than on Arduino due to explicit register configuration, devicetree bindings, and asynchronous APIs, and {\myLevelThree} recovers $+$11.6\% on ESP-IDF and $+$14.5\% on Zephyr.
The results clearly show that active hardware interaction via {\labname} boosts pass rates across all models and target hardware platforms.

\begin{figure}[!t]
\centering
\includegraphics[width=0.85\linewidth]{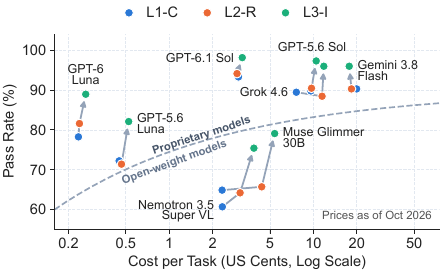}
\vspace{-3.0mm}
\caption{Mean cost per task and pass rate for each model at three feedback levels.
The connected point trajectory for each model follows the agent capability level from {\myLevelOne} to {\myLevelTwo} and {\myLevelThree}.
Costs are calculated from the token counts of each model call using listed API prices.}
\label{fig:acc_vs_cost}
\vspace{-3.0mm}
\end{figure}

\myparatight{Accuracy vs. task cost.}
Fig.~\ref{fig:acc_vs_cost} plots the overall task pass rate against its mean API cost per task (in \$) across three agent capability levels with {\labname}.
Across all eight models, moving from {\myLevelOne} and {\myLevelTwo} to {\myLevelThree} shifts the accuracy vs. cost Pareto frontier sharply upward with only a modest increase in token cost.
It is also interesting to note that there is a slight cost reduction on Gemini~3.8~Flash, where interactive testing replaces long file-exploration loops when using this specific model.
Notably, equipping lightweight models with {\myLevelThree} allows them to match or rival much more expensive models running without interactive hardware feedback: GPT-6~Luna with {\myLevelThree} reaches an 89.0\% pass rate at \$0.0026 per task, matching the {\myLevelOne} accuracy of Grok~4.6 (89.5\% at \$0.075) and Gemini~3.8~Flash (90.3\% at \$0.19) at roughly {30--70}$\times$ lower cost.
At the frontier, GPT-6.1~Sol with {\myLevelThree} achieves the highest overall pass rate (98.2\%) at \$0.032 per task, outperforming models that cost $3\times$--$6\times$ more per task.

\begin{figure}[!t]
\centering
\includegraphics[width=0.9\linewidth]{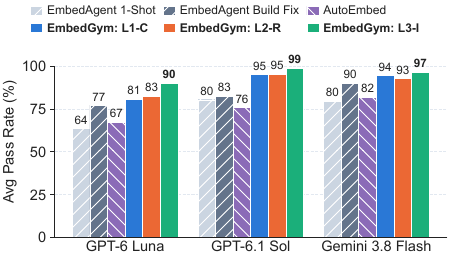}
\vspace{-3.0mm}
\caption{Pass rates of baselines (EmbedAgent~\cite{xu2026embedagent} and AutoEmbed~\cite{yang2026autoembed}) and our agent with {\labnamebf}, evaluated on {\benchnamebf} using the same models, averaged over the four platforms using the Arduino framework.}
\label{fig:baseline_comparison}
\vspace{-3.0mm}
\end{figure}

\myparatight{Comparison with baselines.}
Fig.~\ref{fig:baseline_comparison} compares the agent with {\labname} against EmbedAgent~\cite{xu2026embedagent} and AutoEmbed~\cite{yang2026autoembed} on the four hardware platforms with the Arduino framework (see Table~\ref{tab:eval_leaderboard}) using {\benchname}.
EmbedAgent's single-turn (\mycode{1-Shot}) and one-round compiler repair (\mycode{Build-Fix}) perform well on AVR and ESP32-S3 but drop sharply on nRF52840 (34\%--72\%), as one-shot prompts cannot iteratively explore board-specific SDK headers and library APIs.
The agent with {\myLevelOne} lifts the average pass rates from 77\%/83\%/90\% achieved by EmbedAgent with \mycode{Build-Fix} to 81\%/95\%/94\%, and {\myLevelThree}  further reaches 90\%/99\%/97\% overall. 
AutoEmbed~\cite{yang2026autoembed} averages 67\%/76\%/82\%, which is lower than the agent with {\myLevelOne} on every model.
This is because its fixed pipeline restricts workspace exploration, and its passive serial self-check accepts faulty peripheral logic.
Overall, an agent with capabilities enabled by {\labname} outperforms that with only one-shot repair and passive self-check pipelines.

\subsection{Breakdown of {\labnamebf}'s Feedback Levels}
\label{sec:eval_ablation}

To understand why active hardware interaction outperforms build-only and passive-log feedback in \S\ref{sec:eval_main}, we break down the performance gains introduced by {\labname} across task-level outcome transitions, peripheral modalities, and debugging budgets.

\begin{figure}[!t]
\centering
\includegraphics[width=0.99\linewidth]{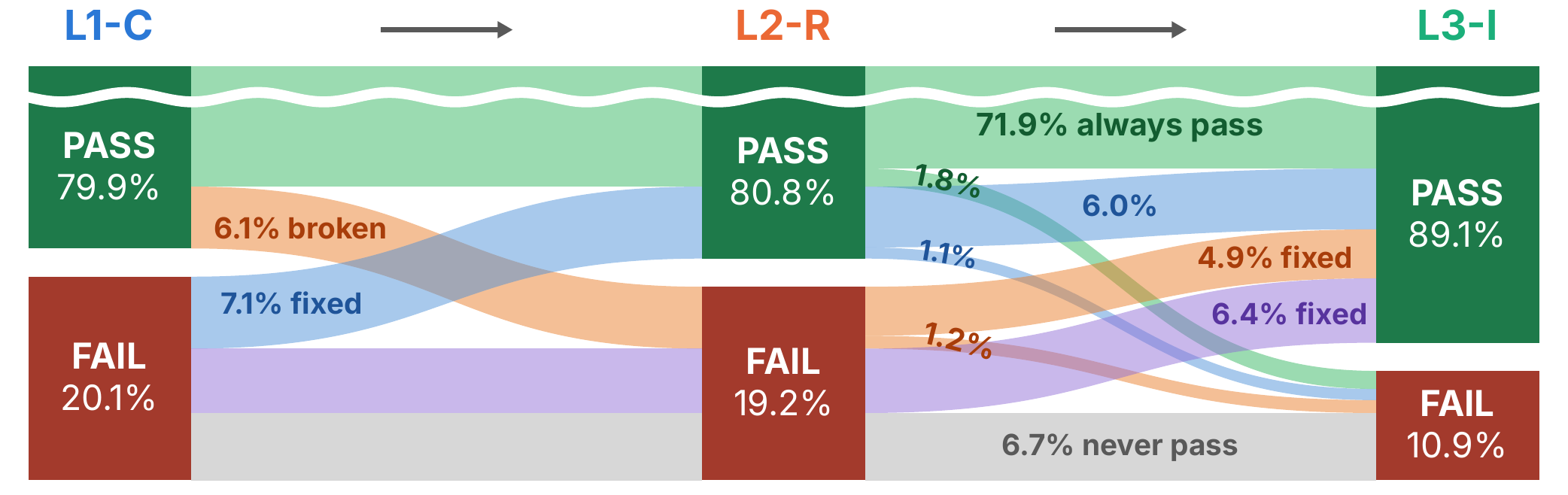}
\vspace{-3.0mm}
\caption{Task success rate with {\labnamebf} from {\myLevelOne}, {\myLevelTwo}, and {\myLevelThree} across 4,784 task-target-model runs; the same color corresponds to the same set of tasks.}
\label{fig:sankey_flow}
\vspace{-3.0mm}
\end{figure}

\myparatight{Task-level outcome transitions.} 
Fig.~\ref{fig:sankey_flow} tracks how individual task outcomes change across all 4,784 runs with {\labname} from {\myLevelOne} to {\myLevelTwo}  and {\myLevelThree}, where each color corresponds to the same set of tasks in {\benchname}.
Moving from {\myLevelOne} to {\myLevelTwo}, 7.1\% of tasks are fixed, but 6.1\% of previously passing tasks are broken at the same time.
This leaves a net pass rate gain of only 0.9\% (79.9\% $\rightarrow$ 80.8\%).
In contrast, moving to {\myLevelThree} breaks only 2.9\% (1.8\% + 1.1\%) of passing tasks, which is less than half the rate in {\myLevelTwo}.
At the same time, {\myLevelThree} fixes 11.3\% of tasks overall, including 4.9\% that were broken in {\myLevelTwo} and 6.4\% that failed in both earlier levels.
Overall, passive serial logs ({\myLevelTwo}) break nearly as many tasks as they fix, whereas active interaction ({\myLevelThree}) fixes far more tasks.

\begin{figure}[!t]
\centering
\includegraphics[width=.75\linewidth]{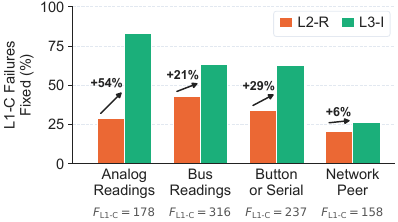}
\vspace{-2.0mm}
\caption{Percentage of {\myLevelOne} failed attempts in each input modality category (Fig.~\ref{fig:bench_arch}(d)) that pass in {\myLevelTwo} or {\myLevelThree}, across all models and hardware targets. $F_{\mathrm{L1\text{-}C}}$ denotes the number of {\myLevelOne} failed attempts in each category.}
\label{fig:modality}
\vspace{-3.0mm}
\end{figure}

\begin{figure}[!t]
\centering
\includegraphics[width=0.98\linewidth]{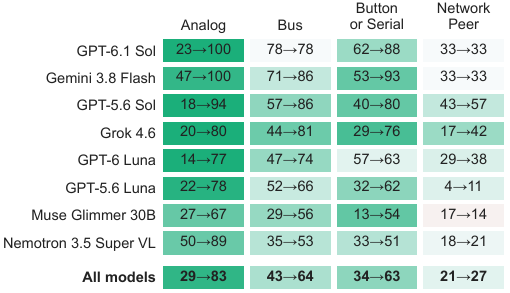}
\vspace{-2.0mm}
\caption{Number of {\myLevelOne} failures that pass at {\myLevelTwo} $\rightarrow$ {\myLevelThree}, by model and input modality category (Fig.~\ref{fig:bench_arch}(d)). Greener cells show larger gains from {\myLevelThree}. Models are ordered by {\myLevelOne} passes; the bottom row combines all models.}
\label{fig:modality_by_model}
\vspace{-3.0mm}
\end{figure}

\myparatight{Breakdown by peripheral modality.} 
In {\myLevelTwo}, without active stimulus, components remain at idle default values, which masks runtime bugs such as ADC scaling, threshold logic, and pin polarity.
Grouping tasks by the six input modality categories from Fig.~\ref{fig:bench_arch}(d) (\S\ref{sec:bench_suite}), Fig.~\ref{fig:modality} shows that actively driving inputs in {\myLevelThree} boosts the percentage of fixed {\myLevelOne} failures over {\myLevelTwo} by $+$54\% on analog sensors, $+$29\% on button or serial inputs, and $+$21\% on bus sensors.
In contrast, network peer tasks show a smaller gain ($+$6\%) because protocol handshakes and server replies already emit diagnostic logs during execution. 
Fig.~\ref{fig:modality_by_model} further breaks down the number of rescued {\myLevelOne} failures from {\myLevelTwo} to {\myLevelThree}  by model, showing that these gains hold across all 8 evaluated models.
The results show that peripherals driven by dynamic physical inputs require active stimulus rather than passive serial logs, further highlighting the benefits of {\myLevelThree} in {\labname}.

\begin{figure}[!t]
\centering
\vspace{-2.0mm}
\includegraphics[width=0.87\linewidth]{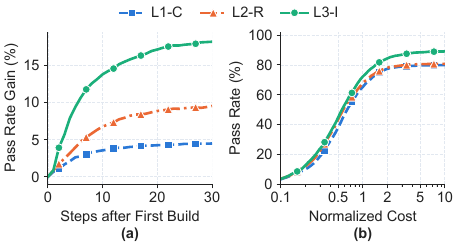}
\vspace{-3.0mm}
\caption{Pass rate scaling across all models and hardware targets: (a)~pass rate gain over each run's first successful build vs.\ post-build steps, and (b)~pass rate vs.\ task cost normalized by each model's mean {\myLevelOne}  cost.}
\label{fig:scaling_roi}
\vspace{-3.0mm}
\end{figure}

\myparatight{Scaling with debugging steps and budget.}
To trace how firmware accuracy evolves across steps for each task, we collect and grade the compiled binary at every step in the agent's trajectory.
Fig.~\ref{fig:scaling_roi} shows that after the first successful build, giving agents more turns without active stimulus yields limited improvement in accuracy. Over 30 additional steps, pass rates rise by only $+$4.5\% in {\myLevelOne} and $+$9.5\% in {\myLevelTwo}, compared to $+$18.2\% in {\myLevelThree} (Fig.~\ref{fig:scaling_roi}(a)).
It can be seen that extra debugging steps and higher budgets only pay off with active hardware interaction.
We further trace how pass rates scale as a function of normalized task budget, and Fig.~\ref{fig:scaling_roi}(b) shows that spending more tokens cannot compensate for missing hardware feedback.
Beyond a normalized cost of 2$\times$, {\myLevelOne} and {\myLevelTwo} plateau near 80\% even at 10$\times$ the initial build cost, whereas {\myLevelThree} already reaches 80\% at 1.5$\times$ cost and continues climbing to 89.1\%.

\subsection{Agent Trajectory and Failure Analysis}
\label{sec:eval_deep_dive}

To explain the accuracy trends shown in \S\ref{sec:eval_main} and \S\ref{sec:eval_ablation}, we inspect the agent trajectories \emph{step by step} to see what actions agents take and when simulator feedback helps or falls short.
We also compare the distinct workflow preferences across models, such as spending turns exploring the SDK, headers, and library examples, versus running compact build-and-test loops.

\begin{figure}[!t]
\centering
\includegraphics[width=0.95\linewidth]{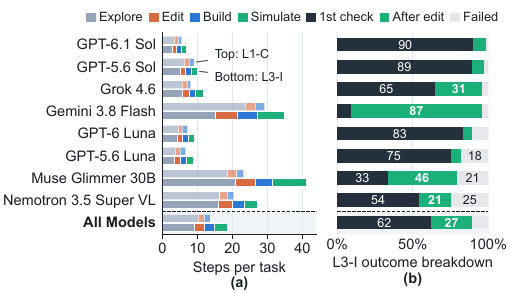}
\vspace{-3.0mm}
\caption{Agent behavior with the simulator. (a) Steps per task by action category in {\myLevelOne} and {\myLevelThree}.
(b) {\myLevelThree} pass rate split into tasks passed on the first simulation check vs. passed after editing firmware based on simulator output.}
\label{fig:agent_behavior}
\vspace{-3.0mm}
\end{figure}

\myparatight{Exploration vs. simulation workflows.} 
To see how agents achieve this gain in {\myLevelThree}, Fig.~\ref{fig:agent_behavior}(a) shows the breakdown of the number of steps each model takes per task into codebase exploration, code editing, building, and simulation.
Pooled across models, average total steps rise from 13.7 in {\myLevelOne} to 18.7 in {\myLevelThree} as agents add 3.9 simulation steps and more edit-and-build cycles.
Models follow two distinct workflow styles: compact agents such as the GPT family and Grok~4.6 spend only 3--6 steps exploring files and add 1.5--2.3 simulation steps on top, whereas exploration-heavy agents like Gemini~3.8~Flash cut file exploration from 23.6 to 15.3 steps and replace it with 8.4 simulation steps and iterative code edits.
Fig.~\ref{fig:agent_behavior}(b) connects these simulator runs to the final 89\% pass rate by separating tasks that already pass on the first simulation check (62\%) from tasks that pass only after the agent edits its code based on simulator output (27\%).
Models with strong initial code, such as GPT-6.1~Sol and GPT-5.6~Sol, pass 89\%--90\% of tasks on the first check and use post-simulation edits to fix the remaining 8\%.
Meanwhile, Grok~4.6, Muse~Glimmer~30B, and Gemini~3.8~Flash pass 31\%, 46\%, and 87\% of all tasks by revising their firmware after seeing simulator output.
Therefore, models balance codebase exploration and simulator testing differently, and editing code after simulation drives the pass rate gain.

\begin{figure}[!t]
\centering
\includegraphics[width=\linewidth]{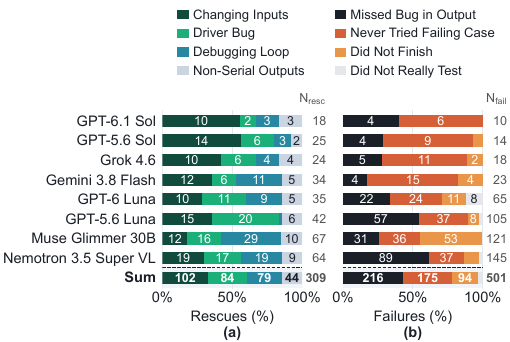}
\caption{(a) Tasks that failed under {\myLevelOne} but passed under {\myLevelThree}, grouped by how the simulator exposed the bug. (b) Tasks that failed under {\myLevelThree}, excluding failures caused by the test setup.
$N_{\text{resc}}$ and $N_{\text{fail}}$ denote the number of tasks per row; models are ordered from fewest to most {\myLevelThree} failures.}
\label{fig:rescue_failure}
\end{figure}

\myparatight{Bug rescue mechanisms and failure causes.} 
To understand what happens inside these simulator checks, we inspect two groups of trajectories in Fig.~\ref{fig:rescue_failure}: the 309 tasks that failed in {\myLevelOne} and were fixed by simulator feedback in {\myLevelThree} (Fig.~\ref{fig:rescue_failure}(a)), and the 501 tasks that still failed in {\myLevelThree}  (Fig.~\ref{fig:rescue_failure}(b)).
Among the 309 fixed tasks, 33.0\% are solved when changing input stimuli exposes reversed threshold or pin polarity logic that stayed hidden at idle.
Another 27.2\% are solved when varied inputs reveal peripheral driver bugs such as wrong register addresses, bitmasks, or scaling formulas.
The rest are fixed through multi-turn debugging loops on runtime errors (25.6\%) or by reading non-serial hardware states such as LCD rows, LEDs, and PWM outputs (14.2\%).
In contrast, the 501 tasks that still failed in {\myLevelThree} show why simulator access alone does not guarantee a fix (Fig.~\ref{fig:rescue_failure}(b)).
Agents rarely fail because they skip testing: in 96.8\% of these failed runs, the agent actively ran the simulator with stimuli.
Instead, 43.1\% of failures happen because the agent overlooked a bug that was already visible in its own simulator output.
Another 34.9\% happen because the agent tested only normal inputs and never tried the failing edge case, while the remaining 18.8\% ran out of steps or budget before finishing.
Overall, active testing fixes input-dependent and driver bugs, while remaining failures come from missed output errors and incomplete test cases.

\subsection{Simulator Fidelity and Workflow Scalability}
\label{sec:eval_fidelity}

We now examine how well simulator grades match physical hardware, how hardware assertions compare with prior benchmarks' validation criteria, and the cost of scaling across targets.

\begin{table}[!t]
\centering
\caption{Sim-to-real hardware-in-the-loop (HIL) grading agreement on physical MCU boards.}
\label{tab:hil_agreement}
\vspace{-3.0mm}
\small
\resizebox{\linewidth}{!}{%
\setlength{\tabcolsep}{4pt}
\begin{tabular}{llcccc}
\toprule
\begin{tabular}[b]{@{}l@{}}\textbf{Physical}\\\textbf{Target}\end{tabular} & \textbf{Emulator} & \textbf{Sampled} & \begin{tabular}[b]{@{}c@{}}\textbf{Sim Pass /}\\\textbf{Real Pass}\end{tabular} & \begin{tabular}[b]{@{}c@{}}\textbf{Sim Fail /}\\\textbf{Real Fail}\end{tabular} & \begin{tabular}[b]{@{}c@{}}\textbf{Agree-}\\\textbf{ment}\end{tabular} \\
\midrule
ATmega328P & \mycode{avr8js} & 10 & 5/5   & 5/5   & 100\% \\
ATmega2560 & \mycode{avr8js} & 10 & 5/5   & 5/5   & 100\% \\
ESP32-S3   & \mycode{qemu}   & 20 & 10/10 & 10/10 & 100\% \\
nRF52840   & \mycode{renode} & 20 & 10/10 & 9/10  & 95.0\% \\
\midrule
\textbf{Total} &                     & \textbf{60} & \textbf{30/30} & \textbf{29/30} & \textbf{98.3\%} \\
\bottomrule
\end{tabular}
\vspace{-2.0mm}
} %
\end{table}

\begin{figure}[!t]
\centering
\includegraphics[width=0.9\linewidth]{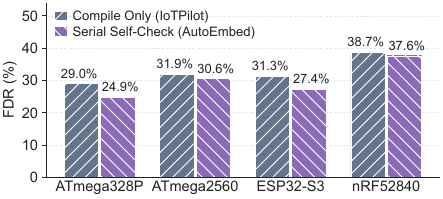}
\caption{False discovery rate (FDR, \%) of prior validation criteria (\emph{Compile Only} in IoTPilot~\cite{gong2025programming} and \emph{Serial Self-Check} in AutoEmbed~\cite{yang2026autoembed}), i.e., the percentage of accepted firmwares that fail {\benchnamebf}'s hardware assertions.}
\label{fig:oracle_fdr}
\end{figure}

\begin{figure*}[!t]
\centering
\includegraphics[width=\linewidth]{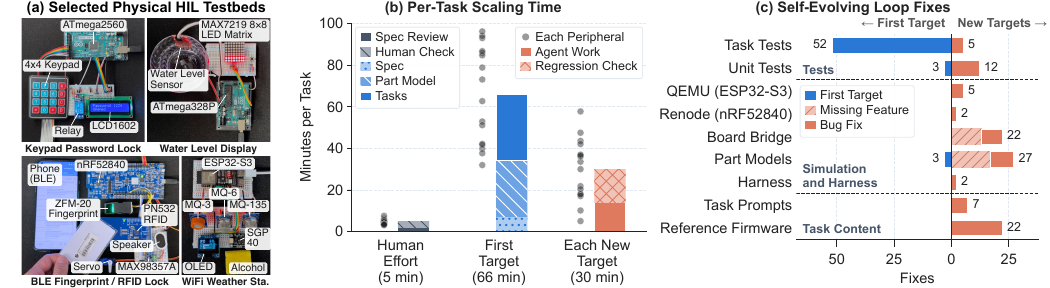}
\caption{Physical hardware verification and workflow scaling evaluation across 16 peripherals: (a) representative physical HIL testbed setups across all four MCU targets (ATmega2560 keypad password lock, ATmega328P water-level display, nRF52840 BLE fingerprint/RFID lock, and ESP32-S3 WiFi multi-sensor station); (b) human and agent time per task for the first board platform and each additional platform, where each dot represents one peripheral; and (c) number of fixes made to each part of the system during verification for the first and additional platforms.}
\label{fig:scaling}
\end{figure*}

\myparatight{Sim-to-real hardware fidelity.}
To check whether simulator grades match real physical hardware, we sample 30 tasks from {\benchname} and pick one passing and one failing firmware per task generated by the agent with {\labname}.
As illustrated in Fig.~\ref{fig:scaling}(a), we wire the physical circuits across all four MCU targets (such as the ATmega2560 keypad password lock, ATmega328P water-level LED matrix display, nRF52840 BLE fingerprint/RFID lock, and ESP32-S3 WiFi multi-sensor station) and flash these 60 binaries onto physical MCU boards.
As shown in Table~\ref{tab:hil_agreement}, behavior on physical hardware matches {\labname}'s emulation output on 59 of the 60 firmwares (98.3\%).
All 30 passing firmwares work properly on real boards, and 29 of the 30 failing firmwares reproduce the exact bug observed in simulation.
The single mismatch is a GPT-6~Luna firmware for the nRF52840 compass-direction task that reads only the $x$- and $y$-axis data registers and never the $z$-axis ones.
Since {\labname} follows the HMC5883L datasheet and keeps the output registers locked until all three axes are read, the heading reading freezes at its first sample and the firmware fails.
The HMC5883L-compatible clone on our physical board does not enforce this lock, which made the faulty firmware appear to work.

\myparatight{Oracle false discovery rate.}
In \S\ref{sec:eval_main}, we showed that compiler output ({\myLevelOne}) and passive serial logs ({\myLevelTwo}) fail to guide agents past runtime hardware bugs, leaving AutoEmbed~\cite{yang2026autoembed} at 67\%--82\% accuracy in Fig.~\ref{fig:baseline_comparison}.
Using our verified hardware assertions as the ground truth, we now show that the same limitation breaks prior benchmarks when they use these checks as their pass criteria (Fig.~\ref{fig:oracle_fdr}).
Under IoTPilot's~\cite{gong2025programming} \emph{Compile Only} criterion, 29.0\%--38.7\% of the firmwares marked as passing across ATmega328P, ATmega2560, ESP32-S3, and nRF52840 actually fail our hardware tests. Even under AutoEmbed's~\cite{yang2026autoembed} \emph{Serial Self-Check}, 24.9\%--37.6\% of accepted firmwares still fail because self-printed debug strings do not reveal bugs when peripheral registers or pins are misconfigured.
Therefore, prior validation methods based on compilation or passive serial logs falsely accept 25\%--39\% of broken firmwares.

\myparatight{Workflow scaling overhead and self-correction.}
Finally, we evaluate our self-evolving workflow end-to-end across 16 peripherals and 26 tasks spanning six hardware interface categories (Fig.~\ref{fig:bench_arch}(c)): three I2C (VL53L0X, SGP40, AMG8833), three UART (PA1616S, TF-Luna, SPS30), three SPI (BME280, MAX7219, ADXL345), three single-wire/1-Wire (DHT11, DS18B20, DS2413), three analog outputs (TMP36, KY-018, KY-023), and one digital GPIO (Relay).
In Stage~1, the reference agent first searches the web for each peripheral and collects 45 datasheets, 34 driver libraries, and 65 module photos across the 16 peripherals.
The agent spends 3.4\,min per peripheral on average.
A human expert then spends an average of 0.98\,min to screen these candidates and narrow down to 19 datasheets, 22 drivers, and 17 photos.
Next, the spec agent takes 5.8\,min per peripheral to synthesize a hardware specification with pin tables and a total of 1{,}085 register, command, and timing checklist rows. 
The human expert spends 1.58\,min per peripheral reviewing the specification and resolving a total of 61 design decisions. 

Across all 16 peripherals, this upfront review (\emph{Spec Review}) takes only 2.0\,min human-time per task before the automated pipeline begins (Fig.~\ref{fig:scaling}(b)).
Once the specification is approved, the agent pipeline runs autonomously.
On the first target (ATmega2560), synthesizing and verifying a peripheral and its tasks takes 66\,min and \$7.94 per task on average (Fig.~\ref{fig:scaling}(b)).
This includes 6.8\,min for Stage~1 reference and spec generation (\emph{Spec}), 29.2\,min for Stage~2 to build and review the peripheral model and unit tests (\emph{Part Model}), and 29.8\,min across Stages~3 and~4 to generate task prompts, circuits, hidden assertions, and reference firmware and to run blind-solve verification (\emph{Tasks} and \emph{Trial Solve}).
In Stage~5, porting each task to the five additional targets takes 30\,min and \$1.97 per task per target, split into 13.9\,min of active agent work (\emph{Agent Work}) and 16.1\,min of full-suite regression runs (\emph{Regression Check}).
At the end, a human wires the physical circuit (as in Fig.~\ref{fig:scaling}(a)), flashes the generated reference firmware onto the real board, and confirms that its behavior matches the task specification (\emph{Human Check}).
This final check takes 2.6\,min per task on average, bringing the total human effort across all stages to 4.6\,min per task.

Fig.~\ref{fig:scaling}(c) shows where the self-evolving loop applies fixes during initial creation versus cross-target porting.
On the first target, almost all fixes land in the newly generated task tests (52 of 58).
These fixes include tightening checks that passed hardcoded or stale outputs, and relaxing overly strict timing windows. During cross-target porting, when task tests are already stable, the verification loop shifts to exposing missing features and bugs across the simulator stack.
It automatically updates peripheral models, board bridges, and low-level timer, GPIO, and bus emulation in \mycode{qemu} and \mycode{renode}.
These results show that our self-evolving workflow scales new peripherals and targets with minimal human supervision while automatically fixing test and simulator defects.

\section{Conclusions}
\label{sec:conclusions}

We presented {\labname}, an interactive virtual hardware environment that lets any coding agent develop and debug embedded firmware without physical hardware, and {\benchname}, a cross-platform benchmark of 598 tasks across 6 MCU-framework pairs and 41 peripherals built on top of it. By coupling instruction-level CPU emulators with behavioral peripheral models, a circuit solver, and virtual WiFi and BLE peers under a deterministic virtual clock, {\labname} allows agents to perturb physical inputs and inspect hardware states during debugging, while {\benchname} grades firmware against hidden test vectors at millisecond precision and scales to new peripherals and targets through a five-stage multi-agent workflow. Our evaluation shows that {\labname} resolves runtime bugs that compilation and passive serial logs miss, raising pass rates across every evaluated model and target, matching physical boards on 98.3\% of sampled firmwares, and letting lightweight models rival frontier ones at a fraction of the cost.

\section*{Acknowledgments}
The work was supported in part by NSF under grants 2112562 and 2625580.

\balance
\bibliographystyle{ACM-Reference-Format}
\bibliography{references}


\begin{thebibliography}{39}


\ifx \showCODEN    \undefined \def \showCODEN     #1{\unskip}     \fi
\ifx \showISBNx    \undefined \def \showISBNx     #1{\unskip}     \fi
\ifx \showISBNxiii \undefined \def \showISBNxiii  #1{\unskip}     \fi
\ifx \showISSN     \undefined \def \showISSN      #1{\unskip}     \fi
\ifx \showLCCN     \undefined \def \showLCCN      #1{\unskip}     \fi
\ifx \shownote     \undefined \def \shownote      #1{#1}          \fi
\ifx \showarticletitle \undefined \def \showarticletitle #1{#1}   \fi
\ifx \showURL      \undefined \def \showURL       {\relax}        \fi
\providecommand\bibfield[2]{#2}
\providecommand\bibinfo[2]{#2}
\providecommand\natexlab[1]{#1}
\providecommand\showeprint[2][]{arXiv:#2}

\bibitem[{Antmicro}(2025)]%
        {renode}
\bibfield{author}{\bibinfo{person}{{Antmicro}}.}
  \bibinfo{year}{2025}\natexlab{}.
\newblock \bibinfo{title}{Renode: Virtual Development Framework for Complex
  Embedded Systems}.
\newblock \bibinfo{howpublished}{\url{https://github.com/renode/renode}}.
\newblock


\bibitem[Austin et~al\mbox{.}(2021)]%
        {austin2021program}
\bibfield{author}{\bibinfo{person}{Jacob Austin}, \bibinfo{person}{Augustus
  Odena}, \bibinfo{person}{Maxwell Nye}, \bibinfo{person}{Maarten Bosma},
  \bibinfo{person}{Henryk Michalewski}, \bibinfo{person}{David Dohan},
  \bibinfo{person}{Ellen Jiang}, \bibinfo{person}{Carrie Cai},
  \bibinfo{person}{Michael Terry}, \bibinfo{person}{Quoc Le}, {et~al\mbox{.}}}
  \bibinfo{year}{2021}\natexlab{}.
\newblock \showarticletitle{Program synthesis with large language models}.
\newblock \bibinfo{journal}{\emph{arXiv preprint arXiv:2108.07732}}
  (\bibinfo{year}{2021}).
\newblock


\bibitem[Bellard et~al\mbox{.}(2005)]%
        {bellard2005qemu}
\bibfield{author}{\bibinfo{person}{Fabrice Bellard} {et~al\mbox{.}}}
  \bibinfo{year}{2005}\natexlab{}.
\newblock \showarticletitle{QEMU, a fast and portable dynamic translator.}
\newblock \bibinfo{journal}{\emph{Usenix ATC, Freenix Track}}
  \bibinfo{volume}{2005} (\bibinfo{year}{2005}), \bibinfo{pages}{41--46}.
\newblock


\bibitem[Chen et~al\mbox{.}(2021)]%
        {chen2021evaluating}
\bibfield{author}{\bibinfo{person}{Mark Chen}, \bibinfo{person}{Jerry Tworek},
  \bibinfo{person}{Heewoo Jun}, \bibinfo{person}{Qiming Yuan},
  \bibinfo{person}{Henrique Ponde De~Oliveira Pinto}, \bibinfo{person}{Jared
  Kaplan}, \bibinfo{person}{Harri Edwards}, \bibinfo{person}{Yuri Burda},
  \bibinfo{person}{Nicholas Joseph}, \bibinfo{person}{Greg Brockman},
  {et~al\mbox{.}}} \bibinfo{year}{2021}\natexlab{}.
\newblock \showarticletitle{Evaluating large language models trained on code}.
\newblock \bibinfo{journal}{\emph{arXiv preprint arXiv:2107.03374}}
  (\bibinfo{year}{2021}).
\newblock


\bibitem[Chen et~al\mbox{.}(2024)]%
        {chen2024teaching}
\bibfield{author}{\bibinfo{person}{Xinyun Chen}, \bibinfo{person}{Maxwell Lin},
  \bibinfo{person}{Nathanael Sch{\"a}rli}, {and} \bibinfo{person}{Denny Zhou}.}
  \bibinfo{year}{2024}\natexlab{}.
\newblock \showarticletitle{Teaching large language models to self-debug}. In
  \bibinfo{booktitle}{\emph{International Conference on Learning
  Representations}}, Vol.~\bibinfo{volume}{2024}. \bibinfo{pages}{8746--8825}.
\newblock


\bibitem[Clements et~al\mbox{.}(2020)]%
        {clements2020halucinator}
\bibfield{author}{\bibinfo{person}{Abraham~A Clements}, \bibinfo{person}{Eric
  Gustafson}, \bibinfo{person}{Tobias Scharnowski}, \bibinfo{person}{Paul
  Grosen}, \bibinfo{person}{David Fritz}, \bibinfo{person}{Christopher
  Kruegel}, \bibinfo{person}{Giovanni Vigna}, \bibinfo{person}{Saurabh Bagchi},
  {and} \bibinfo{person}{Mathias Payer}.} \bibinfo{year}{2020}\natexlab{}.
\newblock \showarticletitle{$\{$HALucinator$\}$: Firmware re-hosting through
  abstraction layer emulation}. In \bibinfo{booktitle}{\emph{29th USENIX
  Security Symposium (USENIX Security 20)}}. \bibinfo{pages}{1201--1218}.
\newblock


\bibitem[{Cognition}(2026)]%
        {frontiercode2026}
\bibfield{author}{\bibinfo{person}{{Cognition}}.}
  \bibinfo{year}{2026}\natexlab{}.
\newblock \bibinfo{title}{{FrontierCode}: Evaluating Software Engineering
  Agents on Real-World Pull Request Mergeability}.
\newblock
  \bibinfo{howpublished}{\url{https://cognition.com/blog/introducing-frontier-code}}.
\newblock


\bibitem[{Cursor}(2026)]%
        {cursorbench2026}
\bibfield{author}{\bibinfo{person}{{Cursor}}.} \bibinfo{year}{2026}\natexlab{}.
\newblock \bibinfo{title}{{CursorBench}: How We Compare Model Quality in
  {Cursor}}.
\newblock \bibinfo{howpublished}{\url{https://cursor.com/blog/cursorbench}}.
\newblock


\bibitem[Fan et~al\mbox{.}(2023)]%
        {fan2023automated}
\bibfield{author}{\bibinfo{person}{Zhiyu Fan}, \bibinfo{person}{Xiang Gao},
  \bibinfo{person}{Martin Mirchev}, \bibinfo{person}{Abhik Roychoudhury}, {and}
  \bibinfo{person}{Shin~Hwei Tan}.} \bibinfo{year}{2023}\natexlab{}.
\newblock \showarticletitle{Automated repair of programs from large language
  models}. In \bibinfo{booktitle}{\emph{2023 IEEE/ACM 45th International
  Conference on Software Engineering (ICSE)}}. IEEE,
  \bibinfo{pages}{1469--1481}.
\newblock


\bibitem[Feng et~al\mbox{.}(2020)]%
        {feng2020p2im}
\bibfield{author}{\bibinfo{person}{Bo Feng}, \bibinfo{person}{Alejandro Mera},
  {and} \bibinfo{person}{Long Lu}.} \bibinfo{year}{2020}\natexlab{}.
\newblock \showarticletitle{$\{$P2IM$\}$: Scalable and hardware-independent
  firmware testing via automatic peripheral interface modeling}. In
  \bibinfo{booktitle}{\emph{29th USENIX Security Symposium (USENIX Security
  20)}}. \bibinfo{pages}{1237--1254}.
\newblock


\bibitem[Gong et~al\mbox{.}(2025)]%
        {gong2025programming}
\bibfield{author}{\bibinfo{person}{Kaijie Gong}, \bibinfo{person}{Wei Dong},
  \bibinfo{person}{Hao Wang}, \bibinfo{person}{Yingqi Peng}, {and}
  \bibinfo{person}{Yi Gao}.} \bibinfo{year}{2025}\natexlab{}.
\newblock \showarticletitle{Programming embedded IoT applications in natural
  language with IoTPilot}. In \bibinfo{booktitle}{\emph{Proceedings of the 23rd
  Annual International Conference on Mobile Systems, Applications and
  Services}}. \bibinfo{pages}{70--82}.
\newblock


\bibitem[Gustafson et~al\mbox{.}(2019)]%
        {gustafson2019toward}
\bibfield{author}{\bibinfo{person}{Eric Gustafson}, \bibinfo{person}{Marius
  Muench}, \bibinfo{person}{Chad Spensky}, \bibinfo{person}{Nilo Redini},
  \bibinfo{person}{Aravind Machiry}, \bibinfo{person}{Yanick Fratantonio},
  \bibinfo{person}{Davide Balzarotti}, \bibinfo{person}{Aur{\'e}lien
  Francillon}, \bibinfo{person}{Yung~Ryn Choe}, \bibinfo{person}{Christophe
  Kruegel}, {et~al\mbox{.}}} \bibinfo{year}{2019}\natexlab{}.
\newblock \showarticletitle{Toward the analysis of embedded firmware through
  automated re-hosting}. In \bibinfo{booktitle}{\emph{22nd International
  Symposium on Research in Attacks, Intrusions and Defenses (RAID 2019)}}.
  \bibinfo{pages}{135--150}.
\newblock


\bibitem[Hendrycks et~al\mbox{.}(2021)]%
        {hendrycks2021measuring}
\bibfield{author}{\bibinfo{person}{Dan Hendrycks}, \bibinfo{person}{Steven
  Basart}, \bibinfo{person}{Saurav Kadavath}, \bibinfo{person}{Mantas Mazeika},
  \bibinfo{person}{Akul Arora}, \bibinfo{person}{Ethan Guo},
  \bibinfo{person}{Collin Burns}, \bibinfo{person}{Samir Puranik},
  \bibinfo{person}{Horace He}, \bibinfo{person}{Dawn Song}, {et~al\mbox{.}}}
  \bibinfo{year}{2021}\natexlab{}.
\newblock \showarticletitle{Measuring coding challenge competence with apps}.
  In \bibinfo{booktitle}{\emph{Proceedings of the 35th Conference on Neural
  Information Processing Systems (NeurIPS 2021) Track on Datasets and
  Benchmarks}}.
\newblock


\bibitem[Jain et~al\mbox{.}(2025)]%
        {jain2025livecodebench}
\bibfield{author}{\bibinfo{person}{Naman Jain}, \bibinfo{person}{Alex Gu},
  \bibinfo{person}{Wen-Ding Li}, \bibinfo{person}{Fanjia Yan},
  \bibinfo{person}{Tianjun Zhang}, \bibinfo{person}{Sida Wang},
  \bibinfo{person}{Armando Solar-Lezama}, \bibinfo{person}{Koushik Sen}, {and}
  \bibinfo{person}{Ion Stoica}.} \bibinfo{year}{2025}\natexlab{}.
\newblock \showarticletitle{Livecodebench: Holistic and contamination free
  evaluation of large language models for code}. In
  \bibinfo{booktitle}{\emph{International Conference on Learning
  Representations}}, Vol.~\bibinfo{volume}{2025}.
  \bibinfo{pages}{58791--58831}.
\newblock


\bibitem[Jimenez et~al\mbox{.}(2024)]%
        {jimenez2024swe}
\bibfield{author}{\bibinfo{person}{Carlos~E Jimenez}, \bibinfo{person}{John
  Yang}, \bibinfo{person}{Alexander Wettig}, \bibinfo{person}{Shunyu Yao},
  \bibinfo{person}{Kexin Pei}, \bibinfo{person}{Ofir Press}, {and}
  \bibinfo{person}{Karthik Narasimhan}.} \bibinfo{year}{2024}\natexlab{}.
\newblock \showarticletitle{Swe-bench: Can language models resolve real-world
  github issues?}. In \bibinfo{booktitle}{\emph{International Conference on
  Learning Representations}}, Vol.~\bibinfo{volume}{2024}.
  \bibinfo{pages}{54107--54157}.
\newblock


\bibitem[Lai et~al\mbox{.}(2023)]%
        {lai2023ds}
\bibfield{author}{\bibinfo{person}{Yuhang Lai}, \bibinfo{person}{Chengxi Li},
  \bibinfo{person}{Yiming Wang}, \bibinfo{person}{Tianyi Zhang},
  \bibinfo{person}{Ruiqi Zhong}, \bibinfo{person}{Luke Zettlemoyer},
  \bibinfo{person}{Wen-tau Yih}, \bibinfo{person}{Daniel Fried},
  \bibinfo{person}{Sida Wang}, {and} \bibinfo{person}{Tao Yu}.}
  \bibinfo{year}{2023}\natexlab{}.
\newblock \showarticletitle{DS-1000: A natural and reliable benchmark for data
  science code generation}. In \bibinfo{booktitle}{\emph{International
  Conference on Machine Learning}}. PMLR, \bibinfo{pages}{18319--18345}.
\newblock


\bibitem[Li et~al\mbox{.}(2026)]%
        {li2026skilled}
\bibfield{author}{\bibinfo{person}{Yiming Li}, \bibinfo{person}{Yuhan Cheng},
  \bibinfo{person}{Mingchen Ma}, \bibinfo{person}{Yihang Zou},
  \bibinfo{person}{Ningyuan Yang}, \bibinfo{person}{Wei Cheng},
  \bibinfo{person}{Hai Li}, \bibinfo{person}{Yiran Chen}, {and}
  \bibinfo{person}{Tingjun Chen}.} \bibinfo{year}{2026}\natexlab{}.
\newblock \showarticletitle{Skilled AI Agents for Embedded and IoT Systems
  Development}. In \bibinfo{booktitle}{\emph{Proceedings of the ACM Conference
  on AI and Agentic Systems}}. \bibinfo{pages}{1248--1252}.
\newblock


\bibitem[Lieret et~al\mbox{.}(2025)]%
        {minisweagent}
\bibfield{author}{\bibinfo{person}{Kilian Lieret}, \bibinfo{person}{Carlos~E.
  Jimenez}, \bibinfo{person}{John Yang}, {and} \bibinfo{person}{{SWE-agent
  Contributors}}.} \bibinfo{year}{2025}\natexlab{}.
\newblock \bibinfo{title}{{mini-swe-agent}}.
\newblock
  \bibinfo{howpublished}{\url{https://github.com/swe-agent/mini-swe-agent}}.
\newblock


\bibitem[Liu et~al\mbox{.}(2023)]%
        {liu2023your}
\bibfield{author}{\bibinfo{person}{Jiawei Liu}, \bibinfo{person}{Chunqiu~Steven
  Xia}, \bibinfo{person}{Yuyao Wang}, {and} \bibinfo{person}{Lingming Zhang}.}
  \bibinfo{year}{2023}\natexlab{}.
\newblock \showarticletitle{Is your code generated by chatgpt really correct?
  rigorous evaluation of large language models for code generation}.
\newblock \bibinfo{journal}{\emph{Advances in neural information processing
  systems}}  \bibinfo{volume}{36} (\bibinfo{year}{2023}),
  \bibinfo{pages}{21558--21572}.
\newblock


\bibitem[Lopes(2025)]%
        {lcgamboa_qemu}
\bibfield{author}{\bibinfo{person}{Luis Claudio~Gamboa Lopes}.}
  \bibinfo{year}{2025}\natexlab{}.
\newblock \bibinfo{title}{{QEMU} Fork with {ESP32} and {ESP32-S3} Emulation
  Support}.
\newblock \bibinfo{howpublished}{\url{https://github.com/lcgamboa/qemu}}.
\newblock


\bibitem[Mera et~al\mbox{.}(2021)]%
        {mera2021dice}
\bibfield{author}{\bibinfo{person}{Alejandro Mera}, \bibinfo{person}{Bo Feng},
  \bibinfo{person}{Long Lu}, {and} \bibinfo{person}{Engin Kirda}.}
  \bibinfo{year}{2021}\natexlab{}.
\newblock \showarticletitle{DICE: Automatic emulation of DMA input channels for
  dynamic firmware analysis}. In \bibinfo{booktitle}{\emph{2021 IEEE Symposium
  on Security and Privacy (SP)}}. IEEE, \bibinfo{pages}{1938--1954}.
\newblock


\bibitem[Merrill et~al\mbox{.}(2026)]%
        {merrill2026terminal}
\bibfield{author}{\bibinfo{person}{Mike Merrill}, \bibinfo{person}{Alexander
  Shaw}, \bibinfo{person}{Nicholas Carlini}, \bibinfo{person}{Boxuan Li},
  \bibinfo{person}{Harsh Raj}, \bibinfo{person}{Ivan Bercovich},
  \bibinfo{person}{Lin Shi}, \bibinfo{person}{Jeong Shin},
  \bibinfo{person}{Thomas Walshe}, \bibinfo{person}{E~Kelly Buchanan},
  {et~al\mbox{.}}} \bibinfo{year}{2026}\natexlab{}.
\newblock \showarticletitle{Terminal-bench: Benchmarking agents on hard,
  realistic tasks in command line interfaces}. In
  \bibinfo{booktitle}{\emph{International Conference on Learning
  Representations}}, Vol.~\bibinfo{volume}{2026}.
  \bibinfo{pages}{40903--40986}.
\newblock


\bibitem[Montero~Crespo(2025)]%
        {velxio}
\bibfield{author}{\bibinfo{person}{David Montero~Crespo}.}
  \bibinfo{year}{2025}\natexlab{}.
\newblock \bibinfo{title}{Velxio: Arduino \& Embedded Board Emulator}.
\newblock
  \bibinfo{howpublished}{\url{https://github.com/davidmonterocrespo24/velxio}}.
\newblock


\bibitem[Ouyang and Srivastava(2024)]%
        {ouyang2024llmsense}
\bibfield{author}{\bibinfo{person}{Xiaomin Ouyang} {and} \bibinfo{person}{Mani
  Srivastava}.} \bibinfo{year}{2024}\natexlab{}.
\newblock \showarticletitle{LLMSense: Harnessing LLMs for high-level reasoning
  over spatiotemporal sensor traces}. In \bibinfo{booktitle}{\emph{2024 IEEE
  3rd Workshop on Machine Learning on Edge in Sensor Systems (SenSys-ML)}}.
  IEEE, \bibinfo{pages}{9--14}.
\newblock


\bibitem[Scharnowski et~al\mbox{.}(2022)]%
        {scharnowski2022fuzzware}
\bibfield{author}{\bibinfo{person}{Tobias Scharnowski}, \bibinfo{person}{Nils
  Bars}, \bibinfo{person}{Moritz Schloegel}, \bibinfo{person}{Eric Gustafson},
  \bibinfo{person}{Marius Muench}, \bibinfo{person}{Giovanni Vigna},
  \bibinfo{person}{Christopher Kruegel}, \bibinfo{person}{Thorsten Holz}, {and}
  \bibinfo{person}{Ali Abbasi}.} \bibinfo{year}{2022}\natexlab{}.
\newblock \showarticletitle{Fuzzware: Using precise $\{$MMIO$\}$ modeling for
  effective firmware fuzzing}. In \bibinfo{booktitle}{\emph{31st USENIX
  Security Symposium (USENIX Security 22)}}. \bibinfo{pages}{1239--1256}.
\newblock


\bibitem[Shaked and {Wokwi Contributors}(2024a)]%
        {avr8js}
\bibfield{author}{\bibinfo{person}{Uri Shaked} {and} \bibinfo{person}{{Wokwi
  Contributors}}.} \bibinfo{year}{2024}\natexlab{a}.
\newblock \bibinfo{title}{{avr8js}: JavaScript {AVR8} Emulator}.
\newblock \bibinfo{howpublished}{\url{https://github.com/wokwi/avr8js}}.
\newblock


\bibitem[Shaked and {Wokwi Contributors}(2024b)]%
        {wokwi}
\bibfield{author}{\bibinfo{person}{Uri Shaked} {and} \bibinfo{person}{{Wokwi
  Contributors}}.} \bibinfo{year}{2024}\natexlab{b}.
\newblock \bibinfo{title}{Wokwi: Online {ESP32}, {STM32}, {Arduino} Simulator}.
\newblock \bibinfo{howpublished}{\url{https://wokwi.com}}.
\newblock


\bibitem[Shinn et~al\mbox{.}(2023)]%
        {shinn2023reflexion}
\bibfield{author}{\bibinfo{person}{Noah Shinn}, \bibinfo{person}{Federico
  Cassano}, \bibinfo{person}{Ashwin Gopinath}, \bibinfo{person}{Karthik
  Narasimhan}, {and} \bibinfo{person}{Shunyu Yao}.}
  \bibinfo{year}{2023}\natexlab{}.
\newblock \showarticletitle{Reflexion: Language agents with verbal
  reinforcement learning}.
\newblock \bibinfo{journal}{\emph{Advances in neural information processing
  systems}}  \bibinfo{volume}{36} (\bibinfo{year}{2023}),
  \bibinfo{pages}{8634--8652}.
\newblock


\bibitem[Wang et~al\mbox{.}(2025)]%
        {wang2025openhands}
\bibfield{author}{\bibinfo{person}{Xingyao Wang}, \bibinfo{person}{Boxuan Li},
  \bibinfo{person}{Yufan Song}, \bibinfo{person}{Frank~F Xu},
  \bibinfo{person}{Xiangru Tang}, \bibinfo{person}{Mingchen Zhuge},
  \bibinfo{person}{Jiayi Pan}, \bibinfo{person}{Yueqi Song},
  \bibinfo{person}{Bowen Li}, \bibinfo{person}{Jaskirat Singh},
  {et~al\mbox{.}}} \bibinfo{year}{2025}\natexlab{}.
\newblock \showarticletitle{Openhands: An open platform for ai software
  developers as generalist agents}. In \bibinfo{booktitle}{\emph{International
  Conference on Learning Representations}}, Vol.~\bibinfo{volume}{2025}.
  \bibinfo{pages}{65882--65919}.
\newblock


\bibitem[Xia et~al\mbox{.}(2025)]%
        {xia2025demystifying}
\bibfield{author}{\bibinfo{person}{Chunqiu~Steven Xia}, \bibinfo{person}{Yinlin
  Deng}, \bibinfo{person}{Soren Dunn}, {and} \bibinfo{person}{Lingming Zhang}.}
  \bibinfo{year}{2025}\natexlab{}.
\newblock \showarticletitle{Demystifying llm-based software engineering
  agents}.
\newblock \bibinfo{journal}{\emph{Proceedings of the ACM on Software
  Engineering}} \bibinfo{volume}{2}, \bibinfo{number}{FSE}
  (\bibinfo{year}{2025}), \bibinfo{pages}{801--824}.
\newblock


\bibitem[Xia and Zhang(2024)]%
        {xia2024automated}
\bibfield{author}{\bibinfo{person}{Chunqiu~Steven Xia} {and}
  \bibinfo{person}{Lingming Zhang}.} \bibinfo{year}{2024}\natexlab{}.
\newblock \showarticletitle{Automated program repair via conversation: Fixing
  162 out of 337 bugs for \$0.42 each using chatgpt}. In
  \bibinfo{booktitle}{\emph{Proceedings of the 33rd ACM SIGSOFT international
  symposium on software testing and analysis}}. \bibinfo{pages}{819--831}.
\newblock


\bibitem[Xu et~al\mbox{.}(2024)]%
        {xu2024penetrative}
\bibfield{author}{\bibinfo{person}{Huatao Xu}, \bibinfo{person}{Liying Han},
  \bibinfo{person}{Qirui Yang}, \bibinfo{person}{Mo Li}, {and}
  \bibinfo{person}{Mani Srivastava}.} \bibinfo{year}{2024}\natexlab{}.
\newblock \showarticletitle{Penetrative ai: Making llms comprehend the physical
  world}. In \bibinfo{booktitle}{\emph{Proceedings of the 25th International
  Workshop on Mobile Computing Systems and Applications}}.
  \bibinfo{pages}{1--7}.
\newblock


\bibitem[Xu et~al\mbox{.}(2026)]%
        {xu2026embedagent}
\bibfield{author}{\bibinfo{person}{Ruiyang Xu}, \bibinfo{person}{Jialun Cao},
  \bibinfo{person}{Mingyuan Wu}, \bibinfo{person}{Wenliang Zhong},
  \bibinfo{person}{Yaojie Lu}, \bibinfo{person}{Ben He},
  \bibinfo{person}{Xianpei Han}, \bibinfo{person}{Shing-Chi Cheung}, {and}
  \bibinfo{person}{Le Sun}.} \bibinfo{year}{2026}\natexlab{}.
\newblock \showarticletitle{Embedagent: Benchmarking large language models in
  embedded system development}. In \bibinfo{booktitle}{\emph{Proceedings of the
  2026 IEEE/ACM 48th International Conference on Software Engineering}}.
  \bibinfo{pages}{630--641}.
\newblock


\bibitem[Yang et~al\mbox{.}(2026)]%
        {yang2026autoembed}
\bibfield{author}{\bibinfo{person}{Huanqi Yang}, \bibinfo{person}{Mingzhe Li},
  \bibinfo{person}{Mingda Han}, \bibinfo{person}{Zhenjiang Li}, {and}
  \bibinfo{person}{Weitao Xu}.} \bibinfo{year}{2026}\natexlab{}.
\newblock \showarticletitle{AutoEmbed: LLM-driven Automated Software
  Development for Generic Embedded IoT Systems}. In
  \bibinfo{booktitle}{\emph{Proceedings of the 2026 ACM/IEEE International
  Conference on Embedded Artificial Intelligence and Sensing Systems}}.
  \bibinfo{pages}{362--376}.
\newblock


\bibitem[Yang et~al\mbox{.}(2024)]%
        {yang2024sweagent}
\bibfield{author}{\bibinfo{person}{John Yang}, \bibinfo{person}{Carlos~E.
  Jimenez}, \bibinfo{person}{Alexander Wettig}, \bibinfo{person}{Kilian
  Lieret}, \bibinfo{person}{Shunyu Yao}, \bibinfo{person}{Karthik Narasimhan},
  {and} \bibinfo{person}{Ofir Press}.} \bibinfo{year}{2024}\natexlab{}.
\newblock \showarticletitle{{SWE-agent}: Agent-Computer Interfaces Enable
  Automated Software Engineering}. In \bibinfo{booktitle}{\emph{Advances in
  Neural Information Processing Systems (NeurIPS)}}.
\newblock


\bibitem[Zaddach et~al\mbox{.}(2014)]%
        {zaddach2014avatar}
\bibfield{author}{\bibinfo{person}{Jonas Zaddach}, \bibinfo{person}{Luca
  Bruno}, \bibinfo{person}{Aurelien Francillon}, \bibinfo{person}{Davide
  Balzarotti}, {et~al\mbox{.}}} \bibinfo{year}{2014}\natexlab{}.
\newblock \showarticletitle{Avatar: A framework to support dynamic security
  analysis of embedded systems' firmwares.}. In
  \bibinfo{booktitle}{\emph{NDSS}}, Vol.~\bibinfo{volume}{14}.
  \bibinfo{pages}{1--16}.
\newblock


\bibitem[Zhang et~al\mbox{.}(2024)]%
        {zhang2024autocoderover}
\bibfield{author}{\bibinfo{person}{Yuntong Zhang}, \bibinfo{person}{Haifeng
  Ruan}, \bibinfo{person}{Zhiyu Fan}, {and} \bibinfo{person}{Abhik
  Roychoudhury}.} \bibinfo{year}{2024}\natexlab{}.
\newblock \showarticletitle{Autocoderover: Autonomous program improvement}. In
  \bibinfo{booktitle}{\emph{Proceedings of the 33rd ACM SIGSOFT International
  Symposium on Software Testing and Analysis}}. \bibinfo{pages}{1592--1604}.
\newblock


\bibitem[Zhou et~al\mbox{.}(2021)]%
        {zhou2021automatic}
\bibfield{author}{\bibinfo{person}{Wei Zhou}, \bibinfo{person}{Le Guan},
  \bibinfo{person}{Peng Liu}, {and} \bibinfo{person}{Yuqing Zhang}.}
  \bibinfo{year}{2021}\natexlab{}.
\newblock \showarticletitle{Automatic firmware emulation through
  invalidity-guided knowledge inference}. In \bibinfo{booktitle}{\emph{30th
  USENIX Security Symposium (USENIX Security 21)}}.
  \bibinfo{pages}{2007--2024}.
\newblock


\bibitem[Zhuo et~al\mbox{.}(2025)]%
        {zhuo2025bigcodebench}
\bibfield{author}{\bibinfo{person}{Terry~Yue Zhuo}, \bibinfo{person}{Minh~Chien
  Vu}, \bibinfo{person}{Jenny Chim}, \bibinfo{person}{Han Hu},
  \bibinfo{person}{Wenhao Yu}, \bibinfo{person}{Ratnadira Widyasari},
  \bibinfo{person}{Imam Nur~Bani Yusuf}, \bibinfo{person}{Haolan Zhan},
  \bibinfo{person}{Junda He}, \bibinfo{person}{Indraneil Paul},
  {et~al\mbox{.}}} \bibinfo{year}{2025}\natexlab{}.
\newblock \showarticletitle{Bigcodebench: Benchmarking code generation with
  diverse function calls and complex instructions}. In
  \bibinfo{booktitle}{\emph{International Conference on Learning
  Representations}}, Vol.~\bibinfo{volume}{2025}.
  \bibinfo{pages}{66602--66656}.
\newblock


\end{thebibliography}

\end{document}